\documentclass[a4paper,11pt]{article}
\pdfoutput=1
\usepackage{jheppub} % for details on the use of the package, please see the JINST-author-manual
\usepackage{lineno}
\usepackage{float}
\newcommand{\muI}{\mu_I}
\newcommand{\muB}{\mu_B}
\newcommand{\LA}{\left \langle}
\newcommand{\RA}{\right \rangle}
\newcommand{\Ns}{N_{\sigma}}
\newcommand{\Nt}{N_{\tau}}
\newcommand{\Tpc}{T_{pc}}

\newcommand{\Ob}{\mathcal{O}}
\newcommand{\M}{\mathcal{M}}
\newcommand{\hs}{\hspace{.6cm}}
\newcommand{\heq}{\hspace{.5mm}}

\newcommand{\csB}{\chi_6^B}

\newcommand{\ceB}{\chi_8^B}

\newcommand{\cnBse}{(\chi_n^B)_2}
\newcommand{\cnBfo}{(\chi_n^B)_4}
\newcommand{\cnBsi}{(\chi_n^B)_6}
\newcommand{\cnBei}{(\chi_n^B)_8}

\newcommand{\ctBT}{(\chi_2^B)_T}

\newcommand{\cfBT}{(\chi_4^B)_T}

\newcommand{\csBT}{(\csB)_T}

\newcommand{\ceBT}{(\ceB)_T}

\newcommand{\csBs}{(\csB)_6}

\newcommand{\cnBT}{(\chi_n^B)_T}
\newcommand{\cnBm}{(\chi_n^B)_m}

\newcommand{\ronBT}{(\rho_1^B)_T}

\newcommand{\rtwBT}{(\rho_2^B)_T}

\newcommand{\rthBT}{(\rho_3^B)_T}

\newcommand{\ronBf}{(\rho_1^B)_4}

\newcommand{\ronBs}{(\rho_1^B)_6}

\newcommand{\rtwBs}{(\rho_2^B)_6}

\newcommand{\rnIT}{(\rho_n^I)_T}

\usepackage{calc}
\newlength{\depthofsumsign}
\newcommand{\nsum}[1][1.2]{% only for \displaystyle
    \mathop{%
        \raisebox
            {-#1\depthofsumsign+1\depthofsumsign}
            {\scalebox
                {#1}
                {$\displaystyle\sum$}%
            }
    }
}

\title{\boldmath A Comparative Analysis between Unbiased Exponential Resummation and Taylor Expansion in Finite-Density QCD with a new phasefactor for Isospin }

\author[a,b]{Sabarnya Mitra}
\affiliation[a]{Centre for High Energy Physics, Indian Institute of Science, Bengaluru 560012, India}
\affiliation[b]{Fak{\"u}ltat F{\"u}r Physik, Universit{\"a}t Bielefeld 33615, Germany}
\emailAdd{smitra@physik.uni-bielefeld.de}
\emailAdd{sabarnyam@iisc.ac.in}

\abstract
{
 The recently introduced unbiased exponential resummation at finite chemical potential has become an important approach which promises to capture reliably the behaviour of higher order conserved charge cumulants appearing otherwise in the finite-density QCD Taylor series of thermodynamic observables. In this paper, we present a thorough analysis of the estimates of charge cumulants upto eighth order and have compared them using Taylor expansion method and unbiased exponential resummation approach for baryon and isospin chemical potentials. We also subsequently compare the different estimates of the radius of convergence obtained using these two methods and check if the zeros of phasefactor for baryochemical potential can indicate something about these estimated values. We propose a new method of finding a non-trivial phasefactor for isospin chemical potential and we attempt explaining the different estimates of radius of convergence from the zeroes of this newly constructed gauge-ensemble average phasefactor for isospin chemical potential. Lastly, we also illustrate kurtosis plots describing the behaviour of overlap problem in isospin chemical potential and check if it maintains consistency with the appearance of zeros of the newly proposed phasefactor.
}

\begin{document}
\maketitle
\flushbottom

\section{Introduction}
\label{sec:Intro}

\hs The strong force, one of the four fundamental forces of Nature is very well described by the quantum field theory of Quantum Chromodynamics (QCD)\,\cite{Gross:2022QCD}. An immensely important and intriguing spectacle in the paradigm of these strong interactions is the QCD phase diagram which features various interesting phases of strongly interacting matter. One of the important aspects of QCD is to explore and map this phase diagram\,\cite{Halasz:1998phasediagram, Rajagopal:1999phasediagram,Stephanov:2006phasediagram,Fukushima:2011phasediagram} as a function of temperature $T$ and baryochemical potential $\muB$. This is pivotal not only for understanding the strong dynamics at various energy scales, but also for illuminating the physics of early universe\,\cite{McGuigan:2008earlyuniverse,Castorina:2015earlyuniverse}. Despite being very robust and seemingly self-explanatory, most of this phase diagram have been constructed out from mere symmetry arguments and analyses of various QCD models. They continue to remain conjectured and await further conclusive evidences. In the quest of such evidences, one often resorts to formulating QCD on a lattice of spacetime\,\cite{Davies:2005latticeQCD,Boyle:2022latticeQCD} mostly because of its remarkable ability to successfully predict results to appreciable degree of precision. Besides offering possible signatures of unexplored phases, the non-perturbative formulation of thermodynamics in lattice QCD also enables one to obtain significant insights about the phase diagram.  Like at present, lattice simulations can well explain the manifestations at finite temperature, zero $\muB$ which resembles the vertical axis of the phase diagram. It also establishes that the phase transition between the hadronic phase and the quark-gluon plasma phases at zero $\muB$ is an analytical crossover\,\cite{Steinbrecher:2018QCDcrossover,Borsanyi:2020QCDcrossover,Bazavov:2018QCDcrossover,Li:2020QCDcrossover,Guenther:2021QCDcrossover}.

However for real finite $\muB$, lattice QCD faces a stumbling block in the shape of sign problem\,\cite{Gupta:2004signproblem,Danzer:2009signproblem,Goy:2016signproblem,Nagata:2021signproblem}. At finite $\muB$, the path integral\,\cite{Palumbo:2002pathintegral} expressing the QCD partition function $Z$ becomes complex with its measure containing a complex fermion determinant\,\cite{Nakamura:2005complexfermiondeterminant} which gives rise to the problem of complex measure. This complex measure hinders implementation of Monte-Carlo importance sampling for estimating this path integral. While reweighting\,\cite{Ejiri:2004reweighting,Li:2006reweighting} this complex measure with a real fermion determinant at zero $\muB$ makes the measure real, the observable part of the integral becomes complex which, after Monte-Carlo estimation provides a phaseangle $\theta$ and a subsequent phasefactor $\cos \theta$ for every gauge configuration of the working ensemble. The severity of this sign problem is governed by the magnitude of this $\cos \theta$ averaged over the entire ensemble, the value of which decreases towards zero for higher values of $\muB$ thereby reflecting increasing severity of the sign problem. This happens because the integrand exhibits tremendous oscillations across positive and negative real values, each of which is also large in magnitude, causing the mean to settle towards zero. This sign problem eventually leads to the breakdown of lattice QCD computations at a finite value of $\muB$ reflected by the non-monotonic behaviour of the calculated observables. This highly restricts our investigation and consequent knowledge of QCD at finite density.

Several new methods\,\cite{Bilic:1987Langevin,Aarts:2016Langevin,Kogut:2019Langevin,Sinclair:2019Langevin,Cristoforetti:2012Thimbles,Cristoforetti:2013Thimbles,Scorzato:2015Thimbles} have been introduced which can successfully avoid this sign problem, most of which unfortunately have very limited applications in QCD explicitly. In the case of QCD, the Taylor expansion around $\muB=0$\,\cite{Gavai:2003Taylor,Ejiri:2003Taylor,Gavai:2004Taylor,Gavai:2008Taylor,Miao:2008Taylor,Falcone:2010Taylor} and analytic continuation of simulations from imaginary to real $\muB$\,\cite{DElia:2002Analytic,Lombardo:2006Analytic,Sakai:2009Analytic} continue to remain the prominent methods for circumventing the sign problem and providing state-of-the-art results for QCD equation of state\,\cite{Fodor:2002EoS,Aoki:2005EoS,Miller:2006EoS,Karsch:2008EoS,Kanaya:2010EoS,Huovinen:2011EoS,Philipsen:2012EoS,Hegde:2014EoS,Bazavov:2017EoS} at finite $\muB$. Resummation approaches like  Pad{\'e}\,\cite{Cvetic:2011Pade,Pasztor:2020Pade,Bollweg:2022Pade} and exponential resummation\,\cite{Mondal:2021exponentialresummation} have been proposed, to improve the slowly convergent Taylor series results. While the former approximates the Taylor coefficients by rational functions where one is interested to find the roots and poles of these functions, the latter provides a direct estimate of $Z$ in the form of an exponential in which the argument comprises finite contributions of lower order Taylor series. Recently, the new formalism of unbiased exponential resummation\,\cite{Mitra:2023unb,Mitra:2022dae} has been introduced for obtaining a more improved QCD Equation of state at finite chemical potential, by obviating the stochastic bias\,\cite{Mitra:2022cumu} and reproducing the exact Taylor series to a given order in $\mu$, where $\mu$ is any generic flavor of chemical potential. This unbiased approach\,\cite{Mitra:2022Bonn} is paramount for recognising the genuine higher order Taylor contributions captured through this approach of resummation.

 Although the breakdown of calculations can be detected by observing the onset of the zeros of phasefactor for $\muB$, it is not the case for $\muI$ since it has no sign problem. Hence, it is not possible to identify a possible breakdown in $\muI$ by looking for the zeros of the phasefactor, which never becomes zero. Although this means that in principle, one can perform unbiased exponential resummation to all real values of $\muI$ extending to infinity, studies suggest that there is a genuine phase diagram\,\cite{deForcrand:2007isospin,Moller:2009isospin,Brandt:2017isospin,Brandt:2018isospin} in the $T-\muI$ plane which illustrates the formation of a pion condensate starting from some finite value of $\muI$ for a low $T$. This signifies that at a low $T$ surely, this formalism is supposed to have a finite radius of convergence in $\muI$ and is expected to experience a breakdown beyond that. Apart from many other objectives, this paper tries to come up with a new indicator for this purpose.

The paper is organised as follows: In Section \ref{sec:Unbiased exp}, we provide a quick overview about the Taylor expansion and the unbiased exponential resummation formalism, which constitute the two main cornerstone of comparisons in this paper. This is accompanied by a brief description about phasefactor of this resummation formalism, and the new idea of a complex phasefactor which is being proposed to identify possible breakdown for $\muI$. In Section \ref{sec:Overlap problem}, we have discussed briefly about the overlap problem and its severity along $\muI$. Starting from the details of scale setting and setup of lattice including random volume sources and gauge configurations used in constructing the Taylor coefficients, the method of error estimation for unbiased exponential resummation formalism are highlighted in Section \ref{sec:setup}. In Section \ref{sec:Results}, we present vivid discussions regarding the comparisons between the sixth and eighth order conserved charge cumulants for both $\muB$ and $\muI$ obtained from Taylor expansion and unbiased exponential resummation. We also demonstrate the same for the different estimates of the radius of convergence, and attempt to observe if the onset of zeros of the gauge ensemble averaged phasefactor $\LA \cos \theta \RA$ and complex phasefactor $\LA e^{i\theta} \RA$ can provide indications consistent with these estimates of radius of convergence for $\muB$ and $\muI$ respectively. Because if this happens, they can then provide consistent indications about the start of breakdown for $\muB$ and $\muI$. In the last part of this section, we illuminate the behaviour of overlap problem in $\muI$ and see if this is consistent with the implications made by the zeros of phasefactor and their onset in $\muI$. We have concluded the paper and its discussion by providing a brief summary in Section \ref{sec:Conclusions}. Throughout this paper, we have used relativistic units ($\hbar = c = 1$) and unit Boltzmann constant, and have often denoted $\mu/T$ as $\mu$. Often we have used this notation $\mu$ in this paper to imply both $\muB$ and $\muI$.

\section{Taylor expansion and Unbiased exponential resummation}
\label{sec:Unbiased exp}

\hs In a $2+1$ flavor QCD with staggered rooted quarks, the grand-canonical partition function $Z$ with suppressed volume dependence for a given temperature $T$ and chemical potential $\mu$ is given as

\begin{equation}
    Z(T,\mu) = \int \mathcal{D}U \heq e^{-S_G\left[T,U\right]} \heq \det \M(T,\mu,U)
    \label{eq:partition function}
\end{equation}
with the fermionic determinant $\det \M(T,\mu,U)$ being 

\begin{equation}
\det \M(T,\mu,U) = \prod_{f=u,d,s} \big[\det \M(T,\mu_f,U)\big]^{1/4} 
\label{eq:staggered fermion}
\end{equation}
 In the above Eqns.\eqref{eq:partition function} and \eqref{eq:staggered fermion}, $U$ represent the gauge field configurations and functional $S_G\left[T,U\right]$ denotes the gluon action. For a thermodynamic system of volume $V$ at temperature $T$, the excess pressure $\Delta P(T,\mu)$ is given as follows:

\begin{equation}
    \frac{\Delta P(T,\mu)}{T^4} = \frac{P(T,\mu) - P(T,0)}{T^4} = \frac{1}{VT^3} \, \ln \left[\frac{Z(T,\mu)}{Z(T,0)}\right]
    \label{eq:excess pressure}
\end{equation}
This measure of excess pressure in Eqn.\eqref{eq:excess pressure}, scaled in powers of $T$ is dimensionless which makes it useful for calculations at finite temperature on a given lattice.

\subsection{Taylor Expansion}

The Taylor Expansion of this excess pressure to $\mathcal{O}(\mu^N)$ is given by 

\begin{equation}
    \frac{\Delta P_N^{\text{T}}(T,\mu)}{T^4} = \sum_{n=1}^{N/2} \frac{\chi_{2n}}{(2n)!} \left(\frac{\mu}{T}\right)^{2n}
    \label{eq:Taylor excess pressure}
\end{equation}
where in the above Eqn.\eqref{eq:Taylor excess pressure}, $\chi_{2n}$ is the conserved charge cumulant of order $2n$. The $n^{th}$ Taylor coefficient is defined as $c_n = \chi_n/n!$. The CP symmetry of QCD instructs this Taylor series of Eqn.\eqref{eq:Taylor excess pressure} to be even in $\mu$ which implies that $N$ is even. In terms of the different correlation functions $D_n$ where the $n^{th}$ order correlation function $D_n$ is defined as 

\begin{equation}
    D_n(T) = \frac{\partial^n \ln \det \M(T,\mu)}{\partial \mu^n}\Bigg|_{\mu=0} 
    \label{eq:correlation functions}
\end{equation}
the first four $\chi_n$ can be expressed as follows: 

\begin{align}
    \chi_1 &= \LA D_1 \RA \notag \\
    \chi_2 &= \LA D_2 \RA + \LA D_1^2 \RA \notag \\
    \chi_3 &= \LA D_3 \RA + 3\,\LA D_2\,D_1 \RA + \LA D_1^3 \RA \notag \\
    \chi_4 &= \LA D_4 \RA + 4\,\LA D_3\,D_1 \RA + 3\,\LA D_2^2 \RA + 6\,\LA D_2\,D_1^2 \RA + \LA D_1^4 \RA 
    \label{eq:Taylor coeff and correlation funs}
\end{align}
Throughout this paper, the notation $\LA \mathcal{O} \RA$ represents Monte-Carlo sampling average of observable $\mathcal{O}$ over all gauge configurations in the ensemble generated at $\mu=0$. All these powers are necessarily the unbiased powers of the respective $D_n$.

\subsection{Unbiased Exponential Resummation}

\subsubsection*{Exponential Resummation}
The method of exponential resummation commences with estimating the partition function directly and then deducing the thermodynamic quantities successively following the subsequent thermodynamic relations. In this approach, the ratio $Z(T,\mu)/Z(T,0)$ in the above Eqn.\eqref{eq:excess pressure} to $\mathcal{O}(\mu^N)$ is given by 
\begin{equation}
    Z_N^{\text{R}}(T,\mu) \equiv \frac{Z(T,\mu)}{Z(T,0)} = \Biggl< \exp \left(\nsum_{n=1}^N \left(\frac{\mu}{T}\right)^n \frac{D_n}{n!}\right)  \Biggr>    
    \label{eq:resummed partition function}
\end{equation}
%
%  
% where the notation $\LA \cdot \RA_0$ in Eqn.\eqref{eq:resummed partition function} represent Monte-Carlo sampling average over the working ensemble of gauge field configurations, each of which has been simulated at $\mu=0$. The expectation value of an observable $\mathcal{O}$ at an arbitrary $\mu$ and temperature $T$ calculated from an ensemble of gauge configurations simulated at $\muS$ at the same $T$ is given by
% \begin{align*}
%     \Big \langle \Ob(\mu) \Big \rangle_{\muS} &= \frac{1}{Z(\muS)} \int \mathcal{D}U \heq e^{-S_G\left[U\right]} \heq \Ob(\mu,U) \heq \det \M(\muS,U)
% \end{align*}
% %
% where we have from Eqn.\eqref{eq:partition function},
% \begin{equation*}
%     Z(\muS) = \int \mathcal{D}U \heq e^{-S_G\left[U\right]} \heq \det \M(\muS,U)
% \end{equation*}
% %
% We have subdued the $T$ dependence in the ongoing discussions of the paper, since the parameter of interest in this work is $\mu$.
%
with the symbols having conventional meanings as explained above.
On an isotropic\footnote{In isotropic lattice, spatial spacing $a_{\sigma} =$ temporal spacing $a_{\tau}$.} lattice of size $N_{\sigma}^3 \cdot N_{\tau}$ in $3+1$ spacetime having $N_{\sigma}$ points in each of the $3$ spatial directions and $N_{\tau}$ points in the temporal direction, the estimate of excess pressure for exponential resummation is obtained as follows:

\begin{equation}
    \frac{\Delta P_N^{\text{R}}(T,\mu)}{T^4} = \left(\frac{N_{\tau}}{N_{\sigma}}\right)^3 \, \ln Z_N^{\text{R}}(T,\mu)
\end{equation}
 where the expression of $Z_N^{\text{R}}$ is given in the above Eqn.\eqref{eq:resummed partition function}. The partition function $Z$ for any real $\mu$ is real-valued by virtue of the CP symmetry of QCD. which makes the $D_n$ given in Eqn.\eqref{eq:correlation functions} real or imaginary for even or odd $n$ respectively. 
 
%  The $D_n$ are the usual temperature dependent $n$-point correlation functions given for every gauge configuration $U$ as the following  

% \begin{equation}
%     D_n(T,U) = \frac{\partial^n \ln \det \M(T,\mu,U)}{\partial (\mu/T)^n}\Bigg|_{\mu=0}
%     \label{eq:derivatives}
% \end{equation}
% 
%\subsection{Effect of finite random volume sources and stochastic bias}
\subsubsection*{Stochastic Bias}
 However since the fermion matrix $\M$ cannot be evaluated exactly using analytical means\,\cite{Ying:1998fermionmatrix}, these correlation functions $D_n$ require numerical estimation. This is done by considering a finite number of random volume sources for every gauge configuration and thereby estimating $D_n$ for every random source for each gauge configuration. Hence in this limit, $D_n$ of Eqn.\eqref{eq:resummed partition function} is replaced by $\overline{D_n}$, which is given as:

\begin{equation}
    \overline{D_n} = \frac{1}{N_R} \nsum_{r=1}^{N_R} D_n^{(r)}
    \label{eq:random volume sources}
\end{equation}
Here in Eqn.\eqref{eq:random volume sources}, $D_n^{(r)}$ is the estimate of $D_n$ in the $r^{th}$ random volume source and $N_R$ is the total number of such random volume sources. One also needs to extract the real part of this complex exponential in Eqn.\eqref{eq:resummed partition function} for determining the estimate of $Z_R^N$ preserving the CP symmetry of QCD. In this limit, the above Eqn.\eqref{eq:resummed partition function} therefore resembles

\begin{equation}
    Z_N^{\text{R}} = \text{Re} \,\LA \left[\exp \left(\nsum_{n=1}^N \left(\frac{\mu}{T}\right)^n \frac{\overline{D_n}}{n!}\right) \right] \RA
    \label{eq:modified resummed partition function}
\end{equation}
where $\overline{D_n}$ is provided in Eqn.\eqref{eq:random volume sources}. Using finite $N_R$ results in stochastic bias in usual exponential resummation formula as well as provides biased estimates of different $D_n$ for every gauge configuration.
 A detailed description about this stochastic bias can be found in Ref.\,\cite{Mitra:2022cumu}. Although this bias decreases as $N_R^{-1}$, it can be very significant depending on the observable probed and the value and order of $\mu$ under consideration. It becomes highly imperative to eliminate this bias which can hinder genuine transparent understanding of underlying physics for finite-density QCD.

\subsection{Unbiased formalism}

\hs The unbiased exponential resummation is formulated to eliminate this stochastic bias. In this formalism, the original structure of the resummation is retained with subtle modification of the argument of exponential. This is done so that on expansion in $\mu$, it produces unbiased estimates of $D_n$ and thereby reproduces Taylor series exactly to a given order in $\mu$. This is true for any flavor of chemical potential, although we have implemented this for $\muB$ and $\muI$ in this paper. For both $\muB$ and $\muI$, the unbiased formalism provides the following expression of the excess pressure $\Delta P/T^4$:

\begin{align}
    \frac{\Delta P_N^{\text{u}}(T,\mu)}{T^4} &= \frac{1}{VT^3} \ln Z_N^{\text{u}}(T,\mu) \hspace{.3cm}\text{where}\notag \\ 
    Z_N^{\text{u}}(T,\mu) &= \text{Re} \,\LA \left[\exp \left(\nsum_{n=1}^N \left(\frac{\mu}{T}\right)^n \frac{\overline{D_n^{\text{u}}}}{n!}\right) \right] \RA
    \label{eq:unbiased exponential resummation}
\end{align}
The above Eqn.\eqref{eq:unbiased exponential resummation} resembles Eqn.\eqref{eq:modified resummed partition function} with the notable exception that in the unbiased formalism, the argument of exponential in the expression of the unbiased partition function $Z_N^{\text{u}}$ comprises the unbiased estimates $\overline{D_n^{\text{u}}}$ of $D_n$ as the coefficient of $\mu^n$. A detailed proof along with other mathematical details of this formalism are presented in \cite{Mitra:2023unb}.

\subsection{Idea of complex isospin chemical potential and associated phasefactor}

The formalism of exponential resummation provides another significant entity which is the phasefactor. The gauge ensemble averaged phasefactor reflects the degree of $\mu$-dependent oscillations of $\det \M$. Close to the zeros of this average phasefactor, the oscillations become highly severe and may often lead to breakdown of calculations, rendering the formalism unreliable. Thus the manifestation of these zeros and their onset are interesting since they can provide a good estimate about the commencement of a possible breakdown. As mentioned regarding Eqn.\eqref{eq:modified resummed partition function}, the exponential of the complex polynomial comprising purely real (imaginary) $D_n$ for even (odd) values of $n$ yield a complex exponential. So for a complex function $z(T,\mu)$, this can be written as $e^z = R\,e^{i\theta}$ where

\begin{align*}
    R\left(T,\mu\right) = \exp{\bigg[\text{Re}\Big(z\left(T,\mu\right)\Big)\bigg]},\, \theta\left(T,\mu\right) = \text{Im}\bigg[z\left(T,\mu\right)\bigg]
\end{align*}

In the case of $\muB$, the exponential is complex for real values of $\muB$ and since one needs to extract the real part of the exponential, the phasefactor in this case is given by $\cos{\theta}$ for every gauge configuration. Hence in this situation we compute and observe the behaviour of $\LA \cos{\theta} \RA$ as a function of $\muB$. 

However for $\muI$, the exponential is always real for real $\muI$ since $D_n$ vanishes for odd $n$. Since $\theta(T,\mu)$\footnote{$\theta(T,\mu) = \sum_{n=1}^N (\frac{\mu}{T})^{2n-1} \text{Im}(D_{2n-1})$} only depends on odd $D_n$, hence $\theta=0$ for every gauge configuration for $\muI$, making $\cos{\theta}=1$. 
This makes identifying the breakdown in case of $\muI$ difficult by inspecting the behaviour of $\LA \cos{\theta} \RA$ for real $\muI$, just like $\muB$. In this paper, we propose to make $\muI$ complex which will make the resummation exponential complex. This will then provide us with a non-trivial phasefactor which will vary for various $\muI$ spanning the complex $\muI$ plane. Since CP symmetry does not guarantee partition function to be real for complex values of chemical potential, we observe the behaviour of the entire complex phasefactor $\LA e^{i\theta} \RA$ for complex $\muI$ as opposed to observing just $\LA \text{Re}(e^{i\theta}) \RA = \LA \cos{\theta} \RA$ for real $\muB$. The two dimensional phasefactor plots for $\muI$ have been constructed in this paper, by plotting the absolute value of $\LA e^{i\theta} \RA$ i.e. $\left|\LA e^{i\theta} \RA\right|$ as a function of  $\left|\muI\right|$.

\section{Overlap problem and its severity}
\label{sec:Overlap problem}

\hs In this section, we give a brief introduction to the overlap problem that becomes predominant in computations involving real $\muI$. Despite the abscence of sign problem in $\muI$, the calculations experience a genuine breakdown beyond a finite radius of convergence $\rho$.

 While generating an ensemble of fermion and gauge field configurations at finite $\muI$ based on extrapolations from the ensemble generated at $\muI=0$, this problem arises
 when the distribution or sample comprising the ratio of fermion determinants at finite $\muI$ to zero $\muI$ i.e. $\det \M(\muI)/\det \M(0)$ becomes heavily tailed. This large tail of the distribution causes Monte-Carlo importance sampling ineffective, and renders reweighting approach inefficient in this limit. One comes across this ratio while reweighting the integrand of the path integral given in Eqn.\,\eqref{eq:partition function}, and this ratio assumes different values for different gauge configuration ensembles. In the realm of reweighting and exponential resummation where $\left|\muI\right| < \rho$, this ratio for a gauge configuration $U$ can be expressed as 

\begin{equation}
    \frac{\det \M(\muI,U)}{\det \M(0, U)} = \exp{\left[\nsum_{n=1}^{\infty} \muI^n \frac{D_n(U)}{n!}\right]}, \hs D_n(U) = \frac{\partial^n }{\partial \muI^n}\ln \det \M(\muI,U)\bigg|_{\muI=0}
    \label{eq:reweighting}
\end{equation}
The volume and temperature dependence of fermion determinant have been suppressed in the above Eqn.\eqref{eq:reweighting}, for the given gauge configuration $U$. These values of the above ratio for a given $\muI$ and for all such $U$ in the ensemble constitute the working sample or distribution and its tail characterises the magnitude and extent of overlap problem. A heavily tailed distribution\footnote{These distributions have large sample variance and often the sample mean is drastically different from the population mean.} will have greater extent of overlap problem. This is because in these distributions, the values radically different from the distribution mean, manifest with appreciable probability and this trait tends to change the sample statistics by a large extent. Although standard deviation can prove to be a reliable estimate characterising the heavy tail, a better quantitative measure is kurtosis $\kappa$ which is the standardised fourth order central moment. For a total of $N$ gauge field configurations, this is represented by:

\begin{equation}
    \kappa(\muI) = \frac{M_4^{\bar{x}}(\muI)}{(\sigma(\muI))^4}
\end{equation}
where $M_4^{\bar{x}}$ is the fourth order central moment and $\sigma$ is the standard deviation of the distribution with mean $\bar{x}$. These are defined as
\begin{align}
    M_4^{\bar{x}} = \frac{1}{N} \nsum_{i=1}^N (x_i-\bar{x})^4, \hspace{5mm} \sigma &= \left[\frac{1}{N} \nsum_{i=1}^N (x_i-\bar{x})^2\right]^{\frac{1}{2}}, \hspace{5mm} \bar{x} = \frac{1}{N} \nsum_{i=1}^N x_i
\end{align}
where $x_i = \left[\det \M(\muI)/\det \M(0)\right]_i$ is the value of the fermion determinant ratio obtained from $i^{th}$ configuration.
%
% given by 
%
% \begin{equation*}
%     x_i = \left[\frac{\det \M(\muI,U)}{\det \M(0, U)}\right]_i
% \end{equation*}
%
The manifestation of this overlap problem however may be different for different formalism which are adopted suitably for subsequent calculations for probing finite density QCD regime.

\section{Setup of calculations}
\label{sec:setup}

\hs In this work, we have extensively used the data generated by the HotQCD collaboration for its ongoing Taylor expansion calculations and charge fluctuations. We discuss the setup and other important relevant details of this data, in this section. 

The QCD action considered for generating the working data of the calculations in this work follows a $2+1$ flavor signature in which the strange quark is $27$ times more massive than the mass degenerate up and down quarks. This action comprises a Symanzik-improved gauge action\,\cite{Symanzik:1983gauge,Symanzik:1983gauge2nd}
and the Highly Improved Staggered Quark (HISQ) fermion action\,\cite{Gabrielli:1990Improvement,MILC:2008HISQ,MILC:2010HISQ}. Gauge field configurations of order $\Ob(10^4$ - $10^6)$ are generated in the temperature range $135$~MeV~$\lesssim~T~\lesssim$~$176$~MeV with $\Nt=8$, $12$ and $16$ and $\Ns=4\Nt$. In this work, an isotropic lattice of size $32^3 \cdot 8$ has been used in Euclidean four spacetime, which is Wick rotated from the usual $3+1$ relativistic Minkowski spacetime. Following the relation $T=(a\Nt)^{-1}$, the temperature for each $\Nt$ is varied by varying the isotropic lattice spacing $a$ through the inverse gauge coupling\footnote{$\beta=6/g^2$ where $g$ is the QCD coupling parameter.} $\beta$. For each $a$, the bare light and strange quark masses $m_l(a)$ and $m_s(a)$ are also tuned so that the pseudo-Goldstone pion and kaon masses produced become equal to the physical pion ($\pi$) and kaon ($K$) masses respectively. This fixes the line of constant physics for the lattice setup under consideration. The scale setting is determined using both the Sommer parameter $r_1$ and the kaon decay constant $f_K$. A complete description of the gauge ensembles and scale setting is provided in Ref.~\,\cite{Bollweg:2021vqf}.

To calculate the charge cumulants for unbiased exponential resummation, the baryon and isospin correlation functions $D_1,\dots,D_6$ are estimated stochastically using $500$ Gaussian volume sources on each gauge configuration. For a detailed derivation of these $D_n$, refer to \cite{Allton:2005Taylor}. The exponential-$\mu$ formalism is used to calculate the first four correlation functions, and further higher derivatives are calculated using the linear-$\mu$ formalism. This is because the ultraviolet divergences remain upto fourth power of $\muB$ or $\muI$ and in order to take care of it, the complete formula of these correlation functions must be considered in which besides the linear trace term, one also takes into account the traces involving non-linear $\mu$ derivatives of fermion matrix $\M$.  Using this data, we have calculated the necessary results which are to be discussed vividly in the next section for $\muB$ and $\muI$ using the methods of Taylor expansion and unbiased exponential resummation. These are computed in the range $0 \leqslant \lvert \mu_{B,I}/T \rvert \leqslant 2.5$, using $100K$ configurations for $\muB$ and $20K$ configurations per temperature for $\muI$. More statistics have been used for $\muB$ over $\muI$ because the sign problem in the former tends to lower the signal-noise ratio. Our results have been obtained on $\Nt=8$ lattices for three temperatures namely at $T \sim 135$, $157$ and $176$ MeV. Besides describing the hadronic, crossover and QGP phases of QCD phase diagram, these temperatures have been chosen carefully as being approximately equal to $\Tpc$ and $\Tpc\pm20$~MeV, where $\Tpc=156.5(1.5)$~MeV is the chiral crossover temperature at zero baryon chemical potential $\muB$ for physical values of bare quark masses\,\cite{Steinbrecher:2018phh}. The same temperatures have been chosen as the working temperatures for $\muI$ also. 

In all these calculations involving Taylor expansion and unbiased exponential resummation, we have considered taking $100$ bootstrap samples of the working gauge configuration ensemble. This bootstrapping algorithm used, is based on a chosen random number generator and is implemented to calculate the errorbars associated with values of the observables, computed in this paper.

\section{Results}
\label{sec:Results}

\subsection{For baryon chemical potential}

In this section, we present a comparative study between the measures of baryon cumulants of sixth and eighth orders obtained using the usual Taylor Expansion and the measures of the same that appear on expansion of the unbiased exponential resummation formula. The Taylor coefficients\footnote{Taylor coefficient is just the charge cumulant scaled by appropriate factorial.} using the Taylor expansion are procured from the different unbiased powers of the corresponding baryon or isospin correlation functions, as given in Eqn.\eqref{eq:Taylor coeff and correlation funs}. 
As mentioned before, the manifestation of the stochastic bias has already been observed in the old exponential resummation formula\,\cite{Mondal:2021exponentialresummation} and not only we have understood the importance of eliminating this bias\,\cite{Mitra:2022cumu}, we also have come up with the unbiased exponential resummation\,\cite{Mitra:2023unb} which replicates the exact Taylor coefficients corresponding to the appropriate powers of $\mu$. Thus within its all-order series, it reproduces Taylor series upto a desired order in $\mu$ which leads to a more improved QCD Equation of state and enabling one to comprehend the true physics of finite-density QCD, to a more reliable extent.

% As mentioned before, the formalism of the unbiased exponential resummation replicates the exact Taylor coefficients corresponding to the appropriate powers of $\mu$ and consequently produces Taylor series upto a desired order in $\mu$, which here refers to both $\muB$ and $\muI$. The manifestation of stochastic bias in the old exponential resummation formula\,\cite{Mondal:2021exponentialresummation} has already been observed and as an outcome, we have understood the importance of eliminating this bias\,\cite{Mitra:2022cumu} in obtaining the correct results thereby knowing the underlying genuine physics of finite-density QCD. 

The higher order charge cumulants or coefficients, are particularly of great significance. Owing to the higher powers of $\mu$ associated with these coefficients, their values strongly influence the behaviour of Taylor series for larger values of $\mu$ ($\mu/T>1$), which is instrumental for understanding finite-density QCD. From a computational point of view, it is very difficult and expensive to evaluate these higher order Taylor coefficients as they require calculating higher order correlation functions and also the higher unbiased powers of the required lower-order correlation functions. So, it is very challenging to evaluate these Taylor coefficients precisely and therefore would be promising if they can be ascertained using some other alternative means, from which they can be obtained with similar precision at the expense of relatively less computational complexities. In this paper, we attempt to testify this using unbiased exponential resummation of second and fourth orders for finite $\muB$ and $\muI$ and compare the subsequent results between these two approaches. 

\begin{figure}[H]
    \centering
    \includegraphics[width=.47\textwidth]{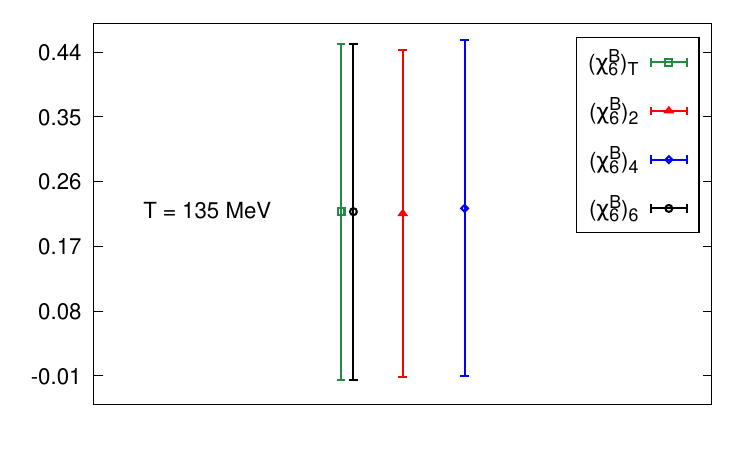}
    \quad
    \includegraphics[width=.47\textwidth]{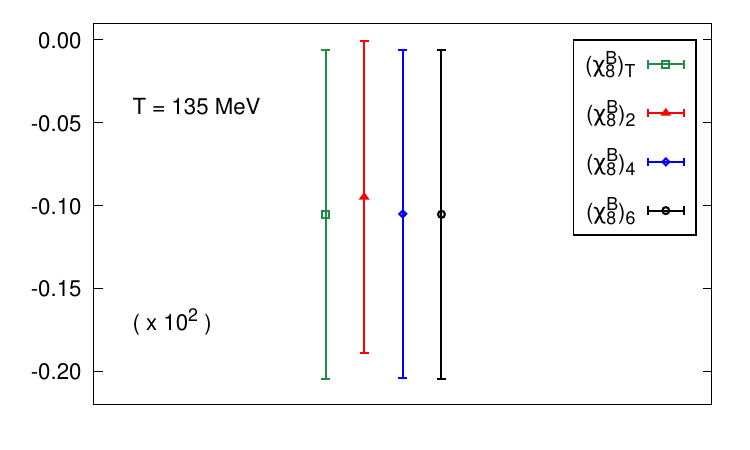}
    \caption{Plots of sixth and eighth order baryon cumulants $\csB$ (left) and $\ceB$ (right) obtained using the Taylor expansion and also from the unbiased exponential resummation of second, fourth and sixth orders at $T=135$ MeV. The green points indicate the usual Taylor estimate of the cumulants whereas the red, blue and black lines represent the same for second, fourth and sixth orders respectively.}
    \label{fig:chi 135_B sixth and eighth}
\end{figure}

Fig.\ref{fig:chi 135_B sixth and eighth} represents the plots of $\chi_6^B$ and $\chi_8^B$ obtained using the usual Taylor Expansion method and the unbiased exponential resummation approach. The $\cnBT$ depicts the estimate of $\chi_n^B$ acquired using Taylor expansion, whereas $\cnBm$ represents the same procured from the unbiased exponential resummation of order $m$, in which correlation functions only upto order $m$ are being taken into account. The subsequent plots in this paper follow this nomenclature of symbols only, even for the plots of $\muI$ in the later section of the paper.
As mentioned before while estimating $\cnBm$, all the correlation functions from $D_1$ to $D_m$ are included and all the higher correlation functions starting from $D_{m+1}$ are ruled out. It is therefore very evident that the sixth order unbiased exponential resummation contains all the necessary unbiased powers of $D_n$ upto $n=6$ and hence in principle, $\csBs =\csBT$ (Eqn.\eqref{eq:Taylor coeff and correlation funs}). This is clearly demonstrated in the left plot of Fig.\ref{fig:chi 135_B sixth and eighth} where the green and the black lines representing respective $\csBT$ and $\csBs$ matches exactly, with the errorbars also in perfect alignment with each other. Strictly speaking, this argument holds true for all the charge cumulants having the order less than the order of the unbiased resummation. In mathematical terms, this implies $(\chi_n^B)_T = (\chi_n^B)_m$ for all $m \geq n$. This is vividly observed from the following Fig.\ref{fig:chi 135_B second and fourth} where we find the measure of $\chi_2$ is exactly identical for all the orders of unbiased resummation and is equal to the corresponding Taylor estimate $\ctBT$. We also observe that as expected, the estimate of $\chi_4$ differs from $\cfBT$ only for second order resummation calculations. They agree exactly with the Taylor result that is, $\cfBT$ from fourth order onwards in terms of both the mean values and errorbars.

\begin{figure}[H]
    \centering
    \includegraphics[width=.449\textwidth]{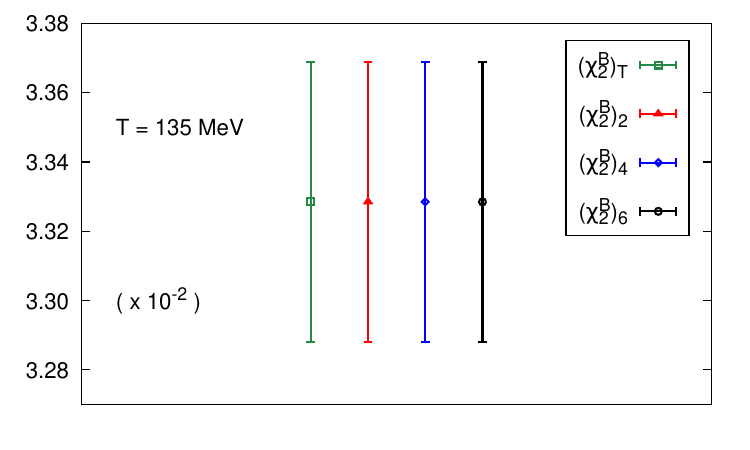}
    \quad
    \includegraphics[width=.449\textwidth]{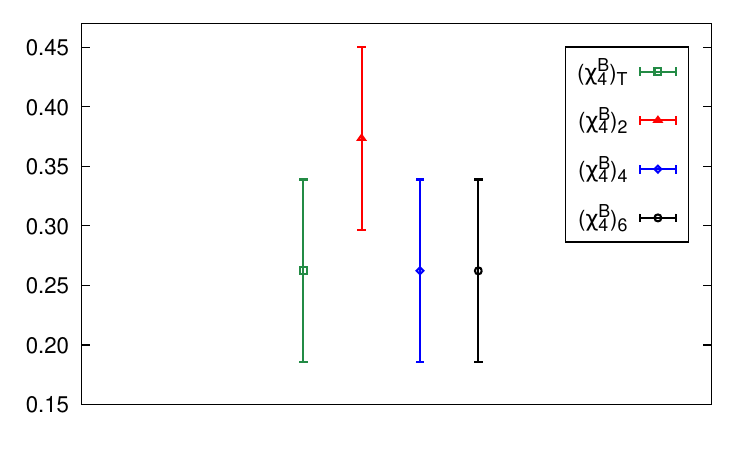}
    \caption{Plots of second and fourth order baryon cumulants $\chi_2^B$ (left) and $\chi_4^B$ (right) obtained using the Taylor expansion and also from the unbiased exponential resummation of second, fourth and sixth orders at $T=135$ MeV. The color and symbol nomenclature remains the same.}
    \label{fig:chi 135_B second and fourth}
\end{figure}

Fig.\ref{fig:chi 135_B sixth and eighth} clearly illustrates that the measurements of $\ceB$ exhibit stochastic fluctuations to a much greater extent, as compared to $\csB$. This is expected as the determination of higher order charge cumulants require evaluating higher order $D_n^B$, more number of diagrams and also higher values of unbiased powers, all of which eventually contribute to the larger magnitudes of these fluctuations. This is very evident from the larger errorbars for $\ceB$ in the right plot compared to $\csB$ in the left plot of Fig.\ref{fig:chi 135_B sixth and eighth}. Although this difference is almost of the order of $10^2$, we find that the resummation estimates align well with the corresponding Taylor estimates for both $\csB$ and $\ceB$ at this temperature. We also observe from this figure that unlike the second order unbiased resummation estimates shown by the red points, the estimates of $\chi_6$ and $\chi_8$ obtained from the fourth and sixth order unbiased resummation (blue and black points) remain more or less in an appreciable agreement with $\csBT$ and $\ceBT$ respectively, where these Taylor results are illustrated by the green points.

% While in Fig.\ref{fig:chi 135_B sixth and eighth} there is a difference between the values and errorbars of $\csBT$ and $\csBt$ as shown by the red points, the blue points representing $\csBf$ clearly reveal that $\csBf$ obtained from the fourth order resummation remains in a very good agreement with $\csBT$, at least within the values with the given order of magnitude This is despite not including the baryon correlation functions $D_5$ and $D_6$ in the estimate of $\csBf$ unlike $\csBT$. 
% This more or less happens to hold also in the case of $\ceB$. As expected, $\chi_8$ measurements exhibit stochastic fluctuations to a much greater extent than $\chi_6$. As shown in Fig.\ref{fig:chi 135_B sixth and eighth} although there is a noticeable shift in the estimate for second order calculations, the estimates obtained from fourth and sixth order resummations seems to be in a very good and commendable agreement with the Taylor estimate at $T=135$ MeV. This is promising, because it seems to indicate that one can only estimate correlation functions upto $D_4^B$ and can rely on adopting a fourth order unbiased exponential resummation approach to evaluate the estimate of $\ceB$ without too much of requiring to evaluate higher correlation functions from $D_5^B$ onwards. As implied by Fig.\ref{fig:chi 135_B sixth and eighth}, the resulting $\ceBt$ obtained would not differ drastically from $\ceBT$, which is the corresponding Taylor series estimate of $\ceB$. 

\begin{figure}[H]
    \centering
    \includegraphics[width=.47\textwidth]{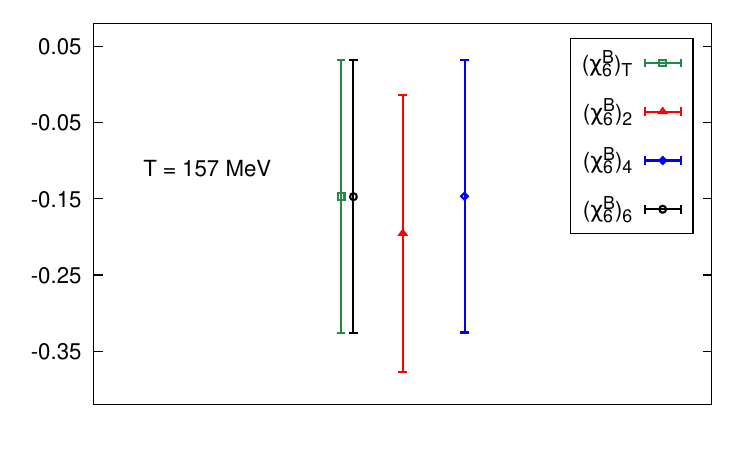}
    \quad
    \includegraphics[width=.47\textwidth]{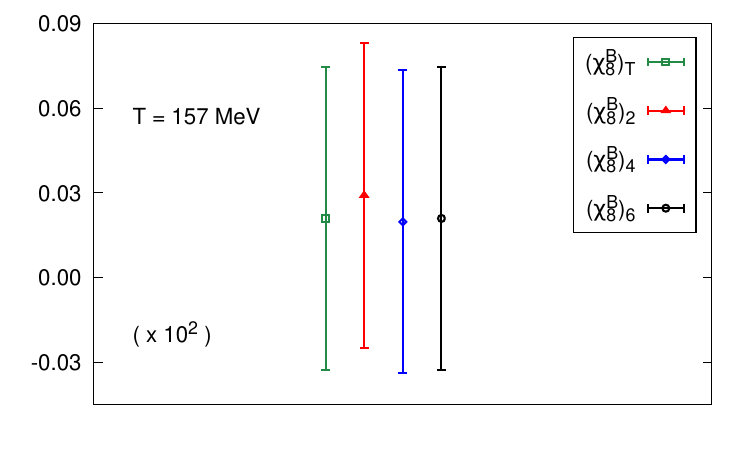} \\
    \includegraphics[width=.47\textwidth]{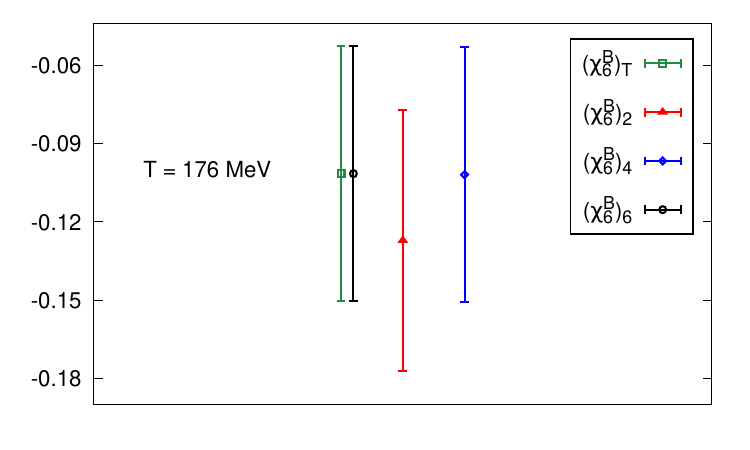}
    \quad
    \includegraphics[width=.47\textwidth]{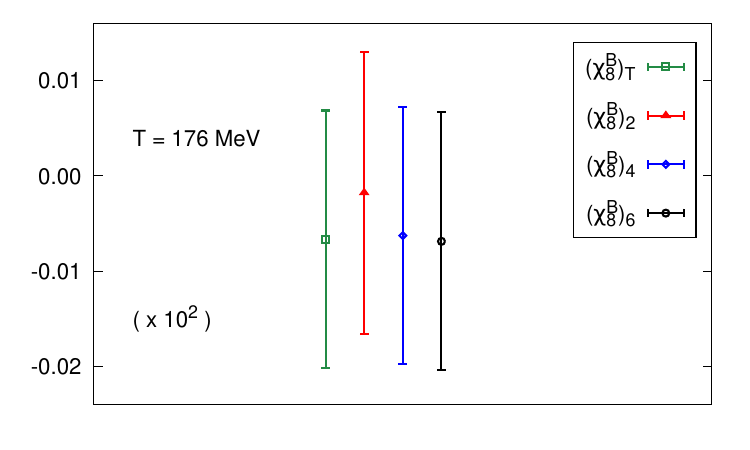}
    
    \caption{Plots of sixth and eighth order charge cumulants $\csB$ (left column) and $\ceB$ (right column) obtained using the Taylor expansion and also from the unbiased exponential resummation of second, fourth and sixth orders. These are obtained at $157$ (top row) and $176$ MeV (bottom row) respectively, with the same meaning of symbols and colors.}
    \label{fig:chi 157_B and 176_B sixth and eighth}
\end{figure}

Similar sort of argument holds true also for the other two temperatures, namely at $157$ and $176$ MeV respectively as shown in the Fig.\ref{fig:chi 157_B and 176_B sixth and eighth}. As compared to Fig.\ref{fig:chi 135_B sixth and eighth}, the difference between the second order resummation estimates and the corresponding Taylor estimates increases for higher temperatures at $157$ and $176$ MeV. This is  demonstrated in Fig.\ref{fig:chi 157_B and 176_B sixth and eighth}, where we find a considerable inequality between $\cnBse$ and $\cnBT$ for $n=6,8$. This may be caused by the increasing magnitude of the remaining higher order terms or diagrams with temperature, which contribute to the Taylor calculation but not in the second order resummation estimate thereby leading to larger differences. Despite this, we observe that the fourth and sixth order resummation estimates, $\cnBsi$ and $\cnBei$ continues to remain in a good agreement with the corresponding Taylor estimates. Although there are some differences, these are much less as compared to $\cnBse$ and we also find for $n=6,8$, the mean values and errorbars $\cnBfo$ and $\cnBsi$ are well-aligned with $\cnBT$ as shown by the blue and black points in Fig.\ref{fig:chi 157_B and 176_B sixth and eighth}.  
  % In both Fig.\ref{fig:chi 135_B sixth and eighth} and Fig.\ref{fig:chi 157_B and 176_B sixth and eighth}, we observe that the second order resummation estimates shift away from   that although the second order resummation estimates $\csBt$ and $\ceBt$ of $\csB$ and $\ceB$ exhibit some noticeable difference from the corresponding Taylor estimates $\csBT$ and $\ceBT$, it is from the fourth order onwards that the resummation estimate values for both $\chi_6$ and $\chi_8$ along with errorbars becomes almost perfectly identical and agrees appreciably well with the corresponding Taylor counterparts. This is found by comparing the blue and black points with the green point in Figs.\ref{fig:chi 135_B sixth and eighth} and \ref{fig:chi 157_B and 176_B sixth and eighth} where the former pair of points represent the fourth and sixth order resummation estimates and the latter illustrates the Taylor estimate. 
 
 It is important to note one interesting feature of this set of observations. Although the second order resummation estimates vary a lot from the Taylor estimates, the difference becomes perceptibly small from fourth order onwards and both the mean values as well as the associated errorbars of the resummation calculations seem to agree well with the Taylor counterparts. This holds true for all the four charge cumulant measurements at all the three working temperatures namely $135$, $157$ and $176$ MeV which represent hadronic, crossover and plasma phases of the QCD phase diagram for physical quark and pion masses. This may seemingly suggest that the calculation of unbiased estimates upto $D_4$ for $\muB$ or equivalently upto $\mathcal{O}(\muB^4)$ and thereafter performing exponential resummation may be good enough to reproduce Taylor series upto $\mathcal{O}(\muB^8)$ irrespective of whether or not, the stochastic bias from $D_5$ is taken care of. The question of whether this remains valid for even higher powers of $\muB$ beyond $\muB^8$ is certainly a very interesting work for the future and if found true, will have huge benefits by reducing computational work and time, to a great extent.

Apart from the estimates of these individual charge cumulants, we also compare and present results regarding the different estimates of the radius of convergence obtained from the Taylor expansion and the unbiased exponential resummation method. The radius of convergence of a Taylor series determines the value of the parameter of expansion ($\mu$ in this case) upto which the series approximation provides reliable results and hence, it is also significant for identifying the possible breakdown of calculations which use Taylor series approximations. As stated before due to the CP symmetry or particle-antiparticle symmetry of QCD, the Taylor series of QCD excess pressure (see Eqn.\eqref{eq:Taylor excess pressure}) is even in $\mu \equiv \muB/\muI$ for which the radius of convergence $\rho$ in principle is given by 

\begin{equation}
    \rho = \lim_{n \to \infty} \rho_n ,\hspace{2mm} \text{where} \hspace{2mm} \rho_n=\sqrt{\frac{c_{2n}}{c_{2n+2}}}
    \label{eq:est of rad of conv}
\end{equation}
In Eqn.\eqref{eq:est of rad of conv}, $c_n = \chi_n/n!$ is the $n^{th}$ order Taylor coefficient appearing Taylor expansion of excess pressure. Although the radius of convergence is in principle, mentioned mostly in the context of Taylor series, we have measured the different estimates of this radius of convergence $\rho$ in the case of unbiased exponential resummation too. This is possible because on expansion of this resummed series in terms of $\mu$, it resembles a Taylor series in $\mu$ whose resulting coefficients differ from the usual Taylor coefficients. This depends on the argument of the exponential of the resummation formula which in turn is dependent on the resummation order. In this work, this has been done by considering the unbiased resummation to second, fourth and sixth orders in $\muB$ and subsequently measuring the appropriate charge cumulants $\chi_n$ of different orders $n$, before determining $\rho$ using Eqn.\eqref{eq:est of rad of conv}.

\begin{figure}[H]
    \centering
    \includegraphics[width=.325\textwidth]{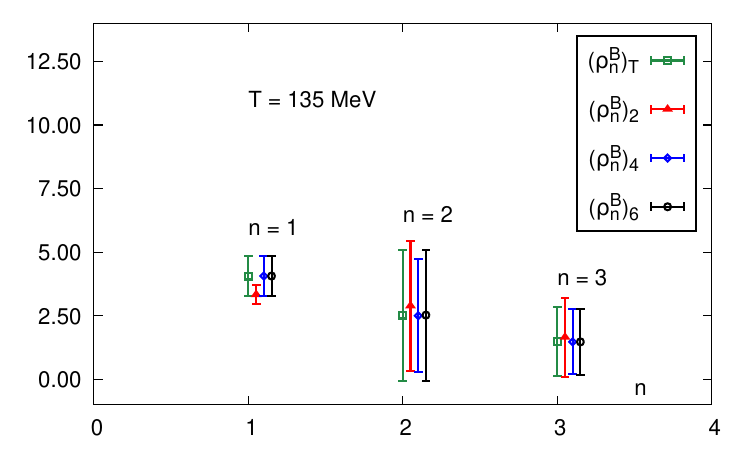}
    \includegraphics[width=.325\textwidth]{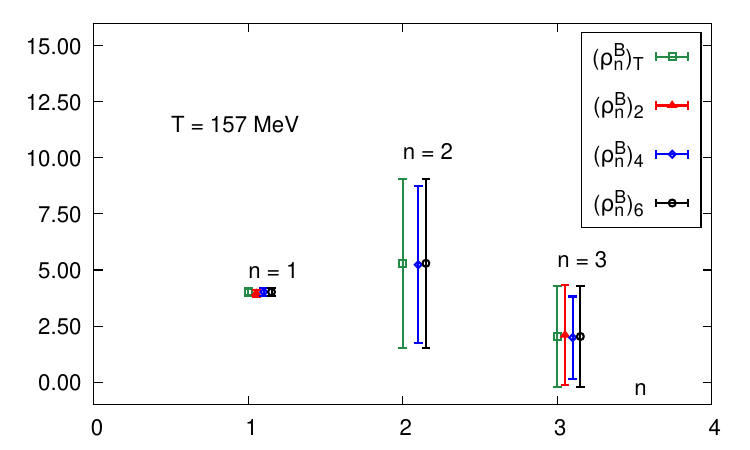} 
    \includegraphics[width=.325\textwidth]{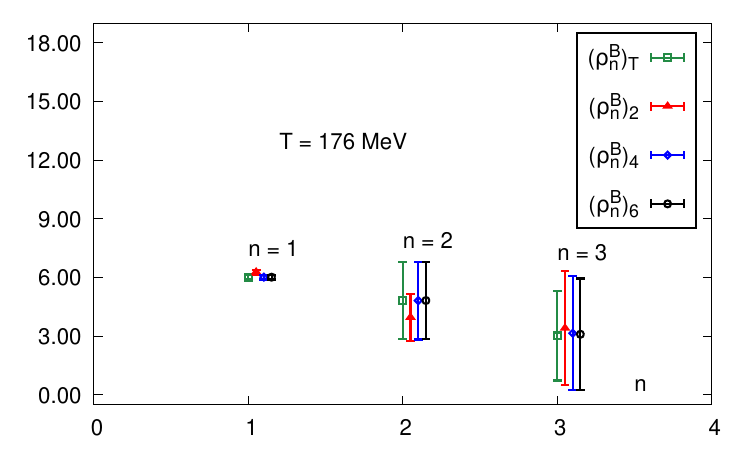}
    
    \caption{Plots of the estimate of $\rho_n=\sqrt{(c_{2n}/c_{2n+2})}$ with $n=1,2,3$ for $\muB$ obtained at $135$ (left), $157$ (center) and $176$ MeV (right). The green line represents the Taylor estimate of the radius of convergence whereas the red, blue and black lines depict the same for unbiased exponential resummation approach upto second, fourth and sixth orders in $\muB$ respectively.}
    \label{fig:all ROC 135_B 157_B and 176_B}
\end{figure}

In this work, we have measured three estimates of the radius of convergence using this unbiased exponential resummation. These estimates are $\rho_1$, $\rho_2$ and $\rho_3$ which are obtained by inserting $n=1,2,3$ in the Eqn.\eqref{eq:est of rad of conv}. 
Since the highest order Taylor series used in this work is of eighth order, the best and most reliable estimate of radius of convergence is $\rho_3 = (c_{6}/c_{8})^{1/2}$. In terms of charge cumulants $\chi$, the three estimates are as follows:

\begin{equation}
    \rho_1 = \sqrt{12\,\,\frac{\chi_{2}}{\chi_{4}}}\hspace{2mm}, \hspace{1cm} \rho_2 = \sqrt{30\,\,\frac{\chi_{4}}{\chi_{6}}}\hspace{2mm}, \hspace{1cm} \rho_3 = \sqrt{56\,\,\frac{\chi_{6}}{\chi_{8}}}
    \label{eq:ROC_from_chi}
\end{equation}

In Fig.\ref{fig:all ROC 135_B 157_B and 176_B}, we have presented plots illuminating these different estimates of $\rho$ as given in Eqn.\eqref{eq:ROC_from_chi}. The different estimates $\rho_1, \rho_2$ and $\rho_3$ of the radius of convergence have been plotted at $135$, $157$ and $176$ MeV respectively. The green points depict the Taylor estimates whereas the red, blue and the black points illustrate the estimates obtained using unbiased exponential resummation of second, fourth and sixth orders respectively. As mentioned before, these estimates have been compared after obtaining them using the Taylor expansion approach and also through the method of unbiased exponential resummation for second, fourth and sixth orders in $\muB$. It is very evident from the previous discussion regarding charge cumulants $\chi_n$ that $\ronBT=(\rho_1^B)_n$ for $n=4,6$. This is because, determination of $\rho_1^B$ requires knowledge of $\chi_2^B$ and $\chi_4^B$, each of which can be completely known to its Taylor estimate from unbiased exponential resummation of fourth order onwards. Similarly in the case of $\rho_2$, one should expect in principle that $\rtwBT = (\rho_2^B)_n$ for $n=6$ only and should not be identical to $(\rho_2^B)_2$ or $(\rho_2^B)_4$. The determination of $\rho_3^B$ is interesting in this case, because in order to calculate $\chi_8^B$ to its Taylor counterpart (see Eqn.\eqref{eq:ROC_from_chi}), one should be needing eighth order unbiased resummation. Being lower than eighth order, all the three orders of unbiased resummation used in this paper are expected to exude some differences over the Taylor estimate $\rthBT$ in the determination of $\rho_3^B$.    

All this set of theoretical expectations are exactly satisfied in Fig.\ref{fig:all ROC 135_B 157_B and 176_B}. It is clearly observed that while all the colored points are differently positioned at $n=3$ for all the three temperatures, the black points merge with the green points for $n=2$ indicating that $\rtwBT=\rtwBs$. Also for $n=1$, both the blue and black points completely overlap with the green points depicting $\ronBT=\ronBf=\ronBs$, and this is true for all the three working temperatures. We clearly find that there is a noticeable distinction between the Taylor estimates shown in green points and the second order unbiased resummation estimates shown by the red points. The Taylor expansion data used has been verified to satisfy that $\rho_1^B$ is minimum and $\rho_2^B$ is maximum at $157$ MeV. 

On a general note, the second order resummation estimates shown in the form of red points exhibit most of the deviation from the corresponding Taylor estimates. This happens for all the three working temperatures, specially for $\rho_1$ and $\rho_2$. Although the second order resummation estimate for $\rho_2$ remains somewhat consistent with the Taylor and the resummation estimates of other orders at $135$ and $176$ MeV, it diverges away completely at $157$ MeV and that is why, it also cannot be captured in the central plot of Fig.\ref{fig:all ROC 135_B 157_B and 176_B}. This happens as $\chi_4^B$ attains a maximum and $\chi_6^B$ attains a minimum at $157$ MeV which is the crossover or the pseudo-critical temperature for the physical values of the quark masses, thereby causing $\rho_2$ to offshoot (see Eqn.\eqref{eq:ROC_from_chi}). Despite this, we also find that most of these deviations happen for the measurements of $\rho_1$ and $\rho_2$ only. While estimating $\rho_3$ which should supposedly be the best estimate of these three estimates, we observe that the resummation estimates of all the second, fourth and sixth orders exhibit consistency and remain in good agreement with one another and the corresponding Taylor estimate. Although there are some differences, these differences are expected as discussed above and even after taking them into account, we clearly observe that they are not dramatically different from what the Taylor results provide us. A detailed study of the values and errorbars would surely be an interesting work for the future.

\begin{figure}[H]
    \centering
    \includegraphics[width=.325\textwidth]{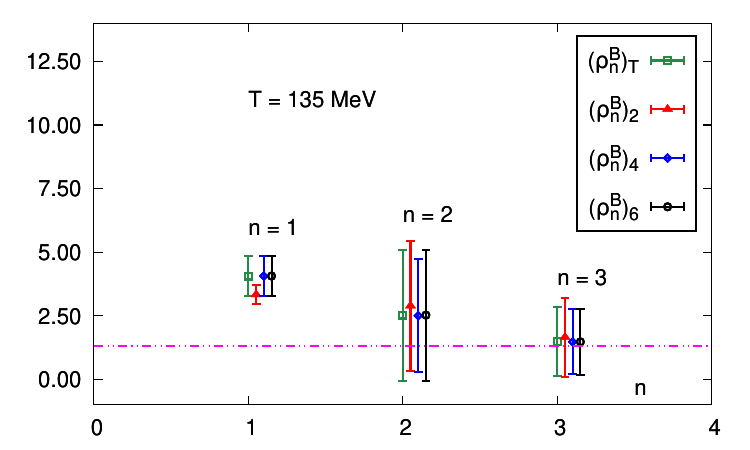}
    \includegraphics[width=.325\textwidth]{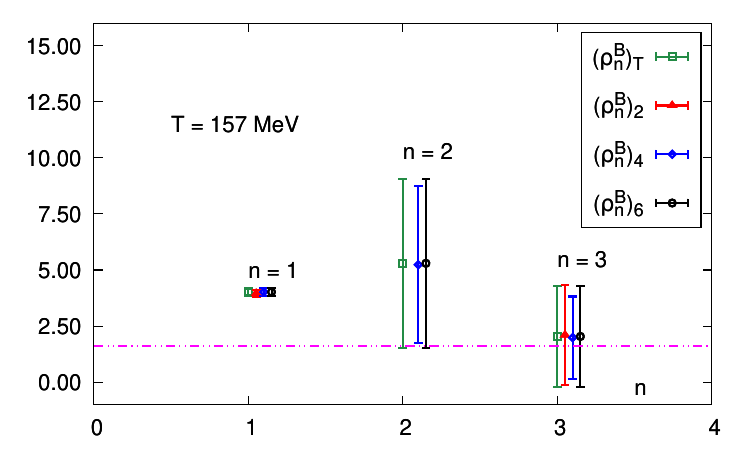}
    \includegraphics[width=.325\textwidth]{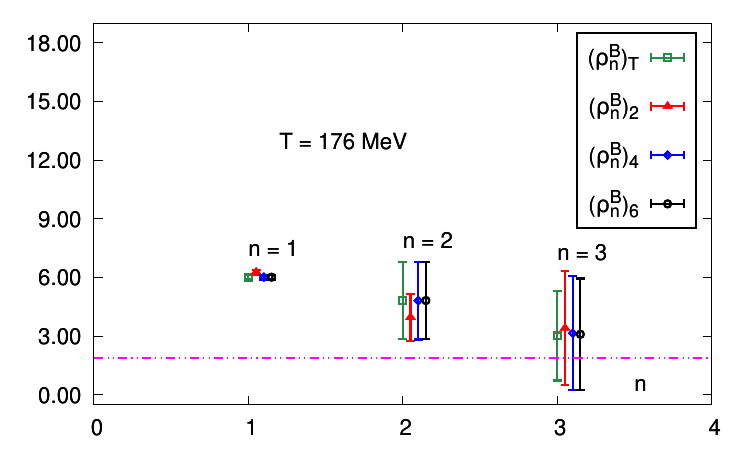}\\
    \includegraphics[width=.325\textwidth]{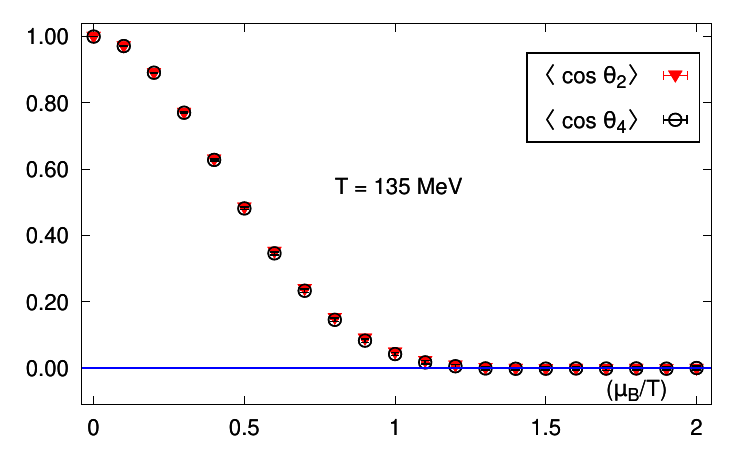}
    \includegraphics[width=.325\textwidth]{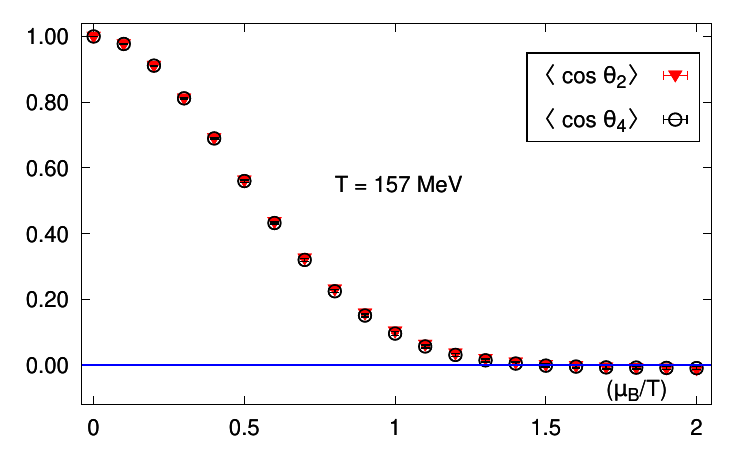} 
    \includegraphics[width=.325\textwidth]{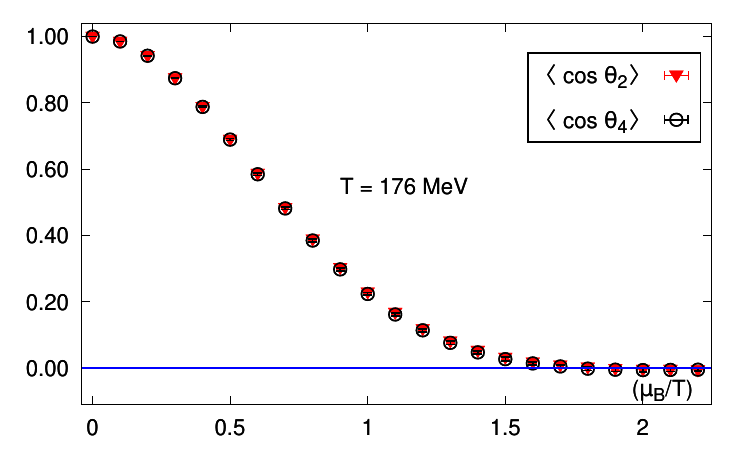}
    
    \caption{(Top row) Plots of $\rho_1^B,\rho_2^B$ and $\rho_3^B$ at $135$ (left), $157$ (middle) and $176$ MeV (right) respectively. (Bottom row) Plots of $\LA \cos \theta \RA$ as a function of $\muB/T$ for second and fourth order unbiased resummation plotted at $135$ (left), $157$ (middle) and $176$ MeV (right) respectively. The red and black points in the phasefactor plots illustrate $\LA \cos \theta \RA$ obtained from unbiased resummation performed upto second and fourth orders in $\muB$ respectively. The blue line in the phasefactor plots indicates the zero line whereas the magenta line in the top row plots illustrate the value of $\muB/T$ from where $\LA \cos \theta \RA=0$.}
    \label{fig:ROC 135,157,176_B all with phasefactor}
\end{figure}

We also present our observations regarding the behaviour of the gauge ensemble averaged phasefactor $\LA \cos \theta \RA$ as a function of $\muB$ in Fig.\ref{fig:ROC 135,157,176_B all with phasefactor}, and subsequently check if the onset of the zeros of this phasefactor coincide with values of estimates of the radius of convergence obtained which  can help identifying the start of the breakdown of calculations. As introduced before, this phasefactor is $e^{i\theta}$ whose real part $\cos \theta$ needs to be analysed for $\muB$ since the partition function is real for real $\muB$. This is obtained from the unbiased exponential resummation and we find that the red and black points in the plots of the lower row of Fig.\ref{fig:ROC 135,157,176_B all with phasefactor} indicate the average phasefactor value procured from unbiased resummation to second order and fourth order respectively. The blue line in the lower row plots is the zero line of the phasefactor, which ascertains the value of $\muB$ from where $\LA \cos {\theta}_n \RA$ becomes zero for $n=2,4$. This value of $\muB$ is illustrated by the magenta dotted line in the upper row plots of the same Fig.\ref{fig:ROC 135,157,176_B all with phasefactor}. From this figure, it is explicitly observed that the zeros of $\LA \cos \theta \RA$ for both the second and fourth orders of unbiased resummation agree very well with each other and also become zero starting from the same value of $\muB$. This remains valid despite this value of $\muB$ being different for different working temperatures. 
Even though the dotted line in magenta does not coincide with the estimates of $\rho_1^B$ for all the temperatures and becomes non-coincident with $\rho_2$ for $T=157$ and $176$ MeV, it coincides well with the estimates of $\rho_3$ for all the three working temperatures at $135$, $157$ and $176$ MeV. This is positive and encouraging, as this implies that the onset of zeros of $\LA \cos \theta \RA$ obtained using unbiased exponential resummation upto fourth order only, can very well indicate the estimated radius of convergence of the Taylor series. This is because we have already discussed that $\rho_3^B$ is the most reliable estimate of the radius of convergence among all these three estimates, and so this demonstrates that the onset of the zeros of $\LA \cos{\theta}\RA$ can provide reliable indications about the estimate of radius of convergence leading to subsequently reliable signs about the onset of breakdown in $\muB$.

\subsection{For isospin chemical potential}

\begin{figure}[H]
    \centering
    \includegraphics[width=.325\textwidth]{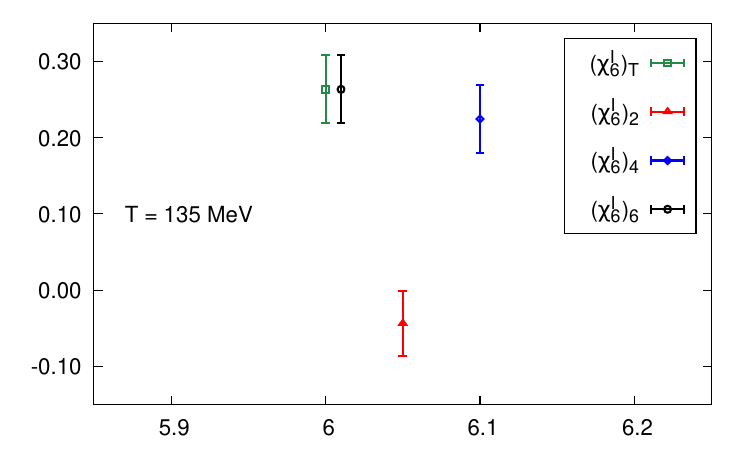}
    \includegraphics[width=.325\textwidth]{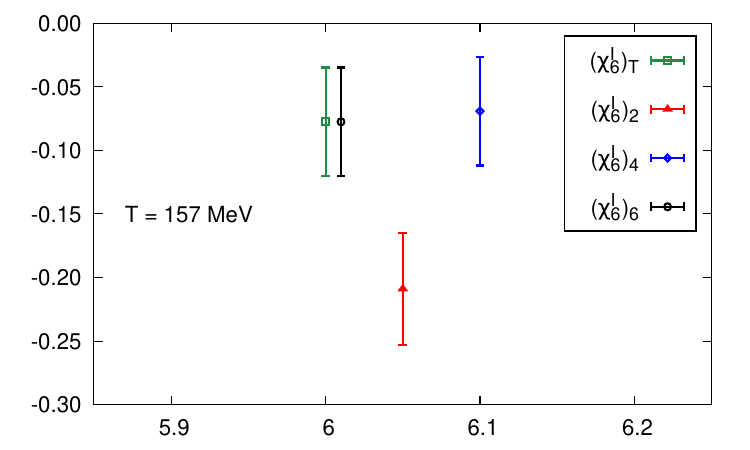}
    \includegraphics[width=.325\textwidth]{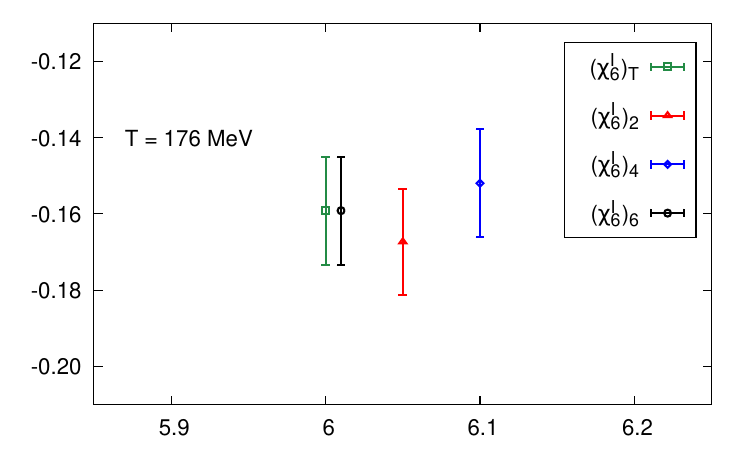} \\
    \hspace{.1mm}
    \includegraphics[width=.325\textwidth]{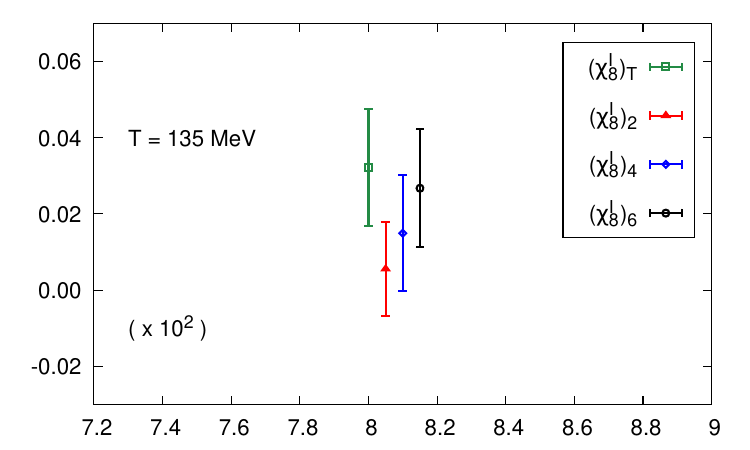}
    \includegraphics[width=.325\textwidth]{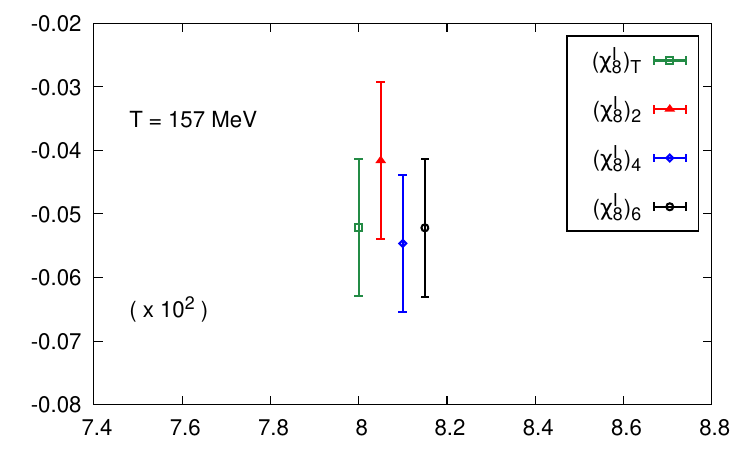}
    \includegraphics[width=.325\textwidth]{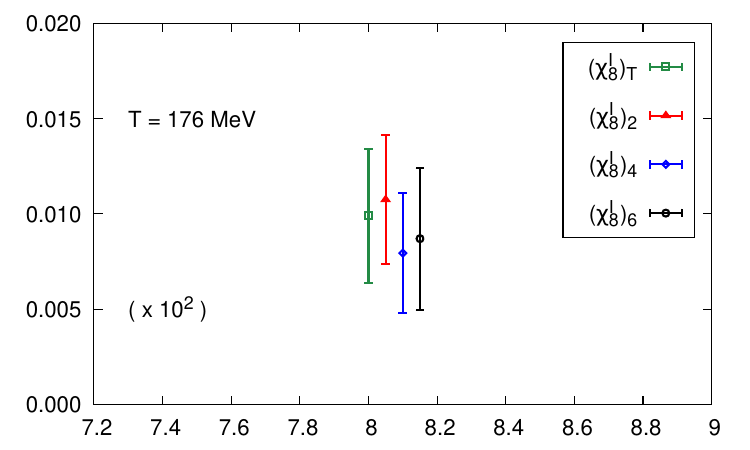}
    
    \caption{Plots for the isospin cumulants $\chi_6^I$ (top row) and $\chi_8^I$ (bottom row) obtained from the methods of Taylor expansions and unbiased exponential resummation. These are obtained for $135$ (left column), $157$ (middle column) and $176$ MeV (right column) respectively with the same color and symbol nomenclature.}
    \label{fig:chi_6 and chi_8 135,157,176_I}
\end{figure}
% \begin{figure}[H]
%     \centering
%     \includegraphics[width=.325\textwidth]{figures/CHI_compar_I_135_c6_Taylor_vs_unb.pdf}
%     \includegraphics[width=.325\textwidth]{figures/CHI_compar_I_157_c6_Taylor_vs_unb.pdf}
%     \includegraphics[width=.325\textwidth]{figures/CHI_compar_I_176_c6_Taylor_vs_unb.pdf} \\
%     \hspace{.1mm}
%     \includegraphics[width=.325\textwidth]{figures/CHI_compar_I_135_c8_Taylor_vs_unb.pdf}
%     \includegraphics[width=.325\textwidth]{figures/CHI_compar_I_157_c8_Taylor_vs_unb.pdf}
%     \includegraphics[width=.325\textwidth]{figures/CHI_compar_I_176_c8_Taylor_vs_unb.pdf}
    
%     \caption{Plots for $\chi_6^I$ (top row) and $\chi_8^I$ (bottom row) obtained from the methods of Taylor expansions and unbiased exponential resummation. These are obtained for $135$ (left col.), $157$ (middle col.) and $176$ MeV (right col.) respectively.}
%     \label{fig:chi_6 and chi_8 135,157,176_I}
% \end{figure}

In this section, we present the same results for isospin chemical potential $\muI$. The comparison regarding $\chi_6$ and $\chi_8$ is illustrated for all the three working temperatures in Fig.\ref{fig:chi_6 and chi_8 135,157,176_I}. 
However as compared to $\muB$ we find from Fig.\ref{fig:chi_6 and chi_8 135,157,176_I}, the resummation estimates differ a lot from the Taylor counterparts. The second order resummation estimate departs the most from the Taylor results whereas the degree of deviation reduces on considering higher orders of unbiased resummation. This difference between $\rnIT$ and $(\rho_n^I)_2$ for $n=6,8$ is maximum at $135$ MeV and reduces with increasing temperature. Unlike $\muB$, we find that the fourth order resummation estimate show some differences with the corresponding Taylor estimate in the determination of $\chi_8^I$.

This may happen because firstly, the odd order isospin correlation functions vanish unlike $\muB$. In case of $\muB$, most of these odd order correlation functions would appear with a negative sign and would tend to nullify the contributions of even order correlation functions coming with a positive sign. An exact reason of this nature of sign still needs to be ascertained. Moreover, we also found that the different even order isospin correlation functions are much greater than baryon correlation functions of similar orders, and this difference increases with temperature and for higher orders and unbiased powers. As a consequence, the contribution of higher order correlation functions $D_6$ and $D_8$ get highly enhanced in case of $\muI$, for which taking unbiased powers only upto $\muI^4$ and performing unbiased exponential resummation is not enough to replicate eighth order Taylor series in $\muI$. In the case of $\muB$, the higher order contributions of $D_6^B$ and $D_8^B$ do not offshoot like $\muI$ and hence, we observe that the agreement between Taylor and resummation estimates become appreciable from fourth order onwards for all the three working temperatures as shown in Fig.\ref{fig:chi 135_B sixth and eighth}.

% Although one resorts to exponential $\mu$ formalism to determine $D_1^I$ to $D_4^I$ and thereby collects additional terms of traces, unlike $\muB$ the resummation estimates do not align well themselves well with the Taylor estimates from the fourth order onwards. This may happen as the odd isospin correlation functions vanish for every gauge configuration and hence 

% However as compared to $\muB$ we find from Fig.\ref{fig:chi_6 and chi_8 135,157,176_I}, the second order resummation estimates for $\chi_6$, namely $\csIt$ vary to a greater extent against the equivalent Taylor estimate $\csIT$. We also find that this difference reduces with increasing temperature for $\chi_6$, seemingly suggesting that the corrections introduced by the unbiased estimates of $D_n$ from $D_3$ to $D_6$, over $D_1$ and $D_2$ in the evaluation of $\csIT$ become less pronounced with increasing temperature. Despite this, we find just like $\muB$ the agreement becomes very good from fourth order onwards and as usual being $\csIs=\csIT$, the black and green points agrees almost completely along with the errorbars. In case of $\chi_8^I$ also, the disagreement is noticeable at second order but again from fourth order the extent of agreement is very good and commendable for all the three working temperatures. As before, this seem to imply that it is important maybe to just consider having unbiased estimates upto $D_4$ only.

\begin{figure}
    \centering
    \includegraphics[width=.325\textwidth]{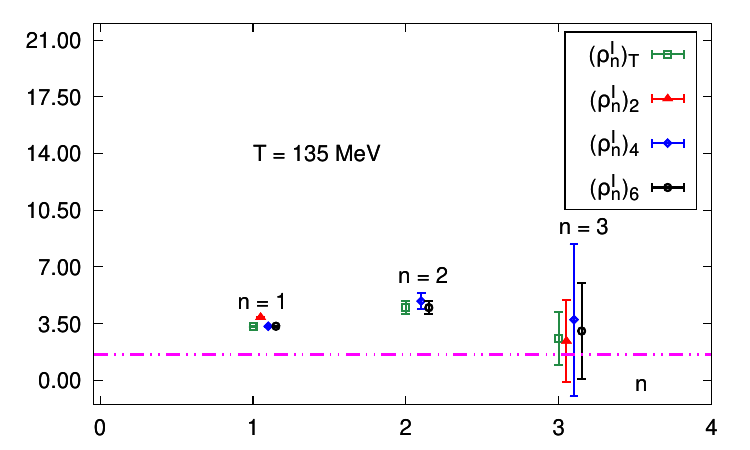}
    \includegraphics[width=.325\textwidth]{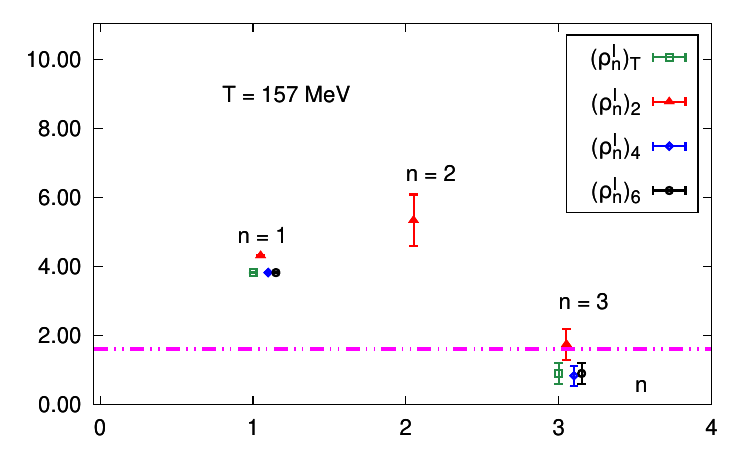}
    \includegraphics[width=.325\textwidth]{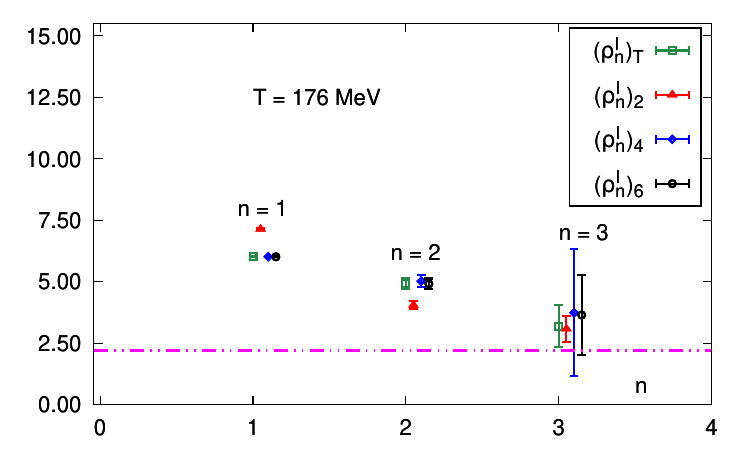} \\
    \hspace{.1mm}
    \includegraphics[width=.325\textwidth]{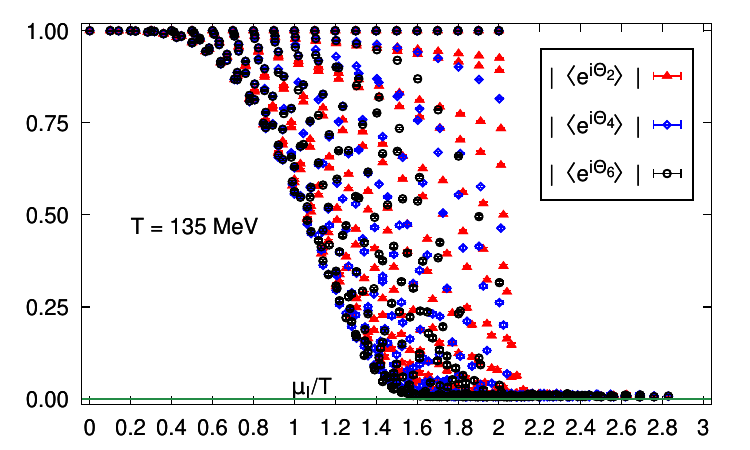}
    \includegraphics[width=.325\textwidth]{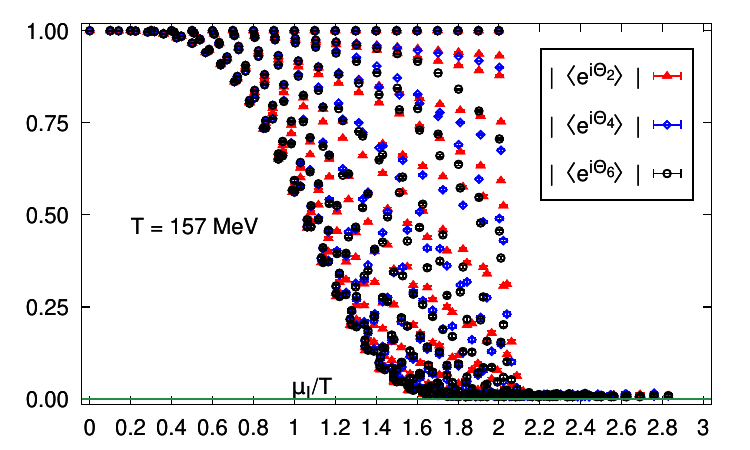}
    \includegraphics[width=.325\textwidth]{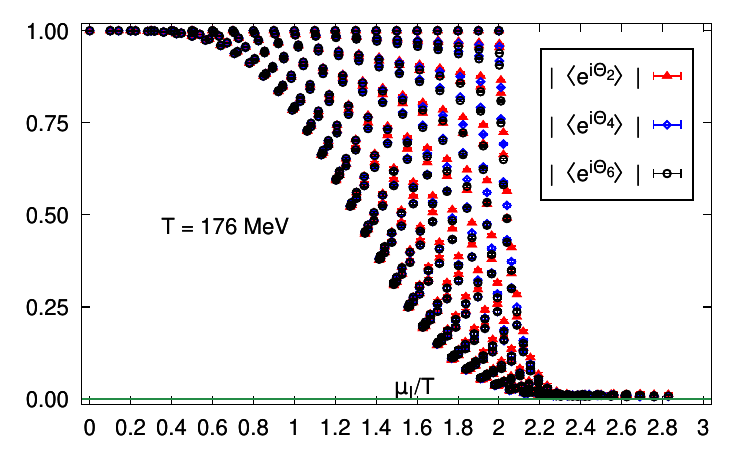}
    
    \caption{(Top row) Plots of $\rho_1^I,\rho_2^I$ and $\rho_3^I$ at $135$ (left), $157$ (middle) and $176$ MeV (right) respectively. (Bottom row) Plots of $\LA e^{i\theta}\RA$ as a function of $\muI/T$ for second, fourth and sixth order unbiased exponential resummation plotted at $135$ (left), $157$ (middle) and $176$ MeV (right) respectively. The red, blue and black points in these bottom row plots depict $\LA e^{i\theta} \RA$ of unbiased resummation to second, fourth and sixth orders in $\muI/T$ respectively, whereas the magenta line in the top row plots illustrate the value of $\muI/T$ manifesting the onset of $\LA e^{i\theta} \RA=0$.}
    \label{fig:ROC 135,157,176_I all with phasefactor_2nd}
\end{figure}

Just like $\muB$, we have also plotted the different estimates of radius of convergence in Fig.\ref{fig:ROC 135,157,176_I all with phasefactor_2nd} for $\muI$ and compared these estimates obtained from the Taylor expansion and unbiased exponential resummation approaches. These different estimates $\rho_1^I$, $\rho_2^I$ and $\rho_3^I$ have been plotted in the upper row of this figure, with the symbols following the nomenclature as mentioned before. Despite having some appreciable differences between the Taylor and unbiased resummation estimates at the level of individual charge cumulants, the estimates of the radius of convergence exhibit some signs of agreement. Although it is observed that the second order resummation estimates shown in red points exhibit appreciable discrepancies with the Taylor results specially at $135$ and $157$ MeV, the estimates obtained from the fourth and sixth orders unbiased resummation comply to a commendable extent and show appreciable consistency with the corresponding Taylor estimates. 

In this paper, we have formulated a new way of evaluating a non-trivial phasefactor for $\muI$ and just like $\muB$, we have checked if this new formulation of the average phasefactor can provide reliable indications about the radius of convergence and possible breakdown in the case of $\muI$. These phasefactor plots have been constructed in the bottom row of this figure. 
Unlike $\muB$, we have determined a complex phasefactor $e^{i\theta}$ in the case of $\muI$. The trick is to make the isospin chemical potential $\muI$ complex and from the consequent complex reweighting factor, compute the gauge ensemble average $\LA e^{i\theta} \RA$ and observe its behaviour for various complex $\muI$. This is because for real $\muI$ the phasefactor is unity for all the gauge configurations due to vanishing odd-order isospin correlation functions and thereby, cannot be used as an indicator for identifying breakdown of calculations. As mentioned before, the partition function can become complex and does not have to be real for complex chemical potentials, and so this allows one to have a non-trivial phasefactor since the phaseangle becomes non-zero and varies with the value of complex $\muI$. 

In this paper, we have traced two-dimensional plots of\,\footnote{$|\LA e^{i\theta} \RA| = \sqrt{\LA \cos \theta \RA^2 + \LA \sin \theta \RA^2}$} $|\LA e^{i\theta} \RA|$ as a function of $\left|\muI\right|$ as demonstrated in the bottom row plots of Fig.\ref{fig:ROC 135,157,176_I all with phasefactor_2nd}. $\left|\muI\right|$ is the radial distance of the complex $\muI$ from the origin in the complex $\muI$ plane. 
This is beneficial in the sense that $\left|\muI\right|$ is identical to $\muI$ along the real axis and it also provides us a new way of exploring the behaviour of phasefactor along real $\muI$. We have also computed the phasefactor from sixth order resummation for $\muI$ since unlike $\muB$, we observed a very appreciable agreement in the estimates of the isospin cumulants $\chi_6^I$ and $\chi_8^I$ between the Taylor results and the sixth order unbiased exponential resummation.
Unlike Fig.\ref{fig:ROC 135,157,176_B all with phasefactor}, there is a possibility of having multiple points of phasefactor for a given $\muI/T$\,\footnote{This is because $x+i\,y$ and $y+i\,x$ will have same value of $\muI/T$, but can have different values of $\LA e^{i\theta} \RA$.} in the phasefactor plots of $\muI$ in Fig.\ref{fig:ROC 135,157,176_I all with phasefactor_2nd} and hence as we observe, there are more number of points as compared to $\muB$. However unlike $\muB$, we find that not all the phasefactor points coincide with the zero phasefactor line at a particular value of $\muI$. Rather we observe that with increasing value of $\muI$, the phasefactor points start going towards the zero phasefactor line and different points coincide with this line at different values of $\muI$, until from a given value all the phasefactor points stack and settle on this zero line, similar to Fig.\ref{fig:ROC 135,157,176_B all with phasefactor} for the case of $\muB$. Despite this, we clearly notice that all the three orders of unbiased resummation do consistently provide the first zeros of $|\LA e^{i\theta} \RA|$ at almost the same value of $\muI$. These first zeros are important and should be noted as these are the very zeros which mark the beginning of increasing fluctuations of fermionic determinant causing $|\LA e^{i\theta} \RA|=0$.  
These values of $\muI$ for $135$, $157$ and $176$ MeV at which these first zeros appear, are illustrated by the dotted magenta line in the upper row plots of Fig.\ref{fig:ROC 135,157,176_I all with phasefactor_2nd}. We need to check if these first zeros can give some sort of indications about the radius of convergence and subsequent breakdown. The phasefactor has been calculated using unbiased resummation upto sixth order and has been plotted for $135$, $157$ and $176$ MeV respectively.

 We clearly sight from Fig.\ref{fig:ROC 135,157,176_I all with phasefactor_2nd} that the magenta line is far from coinciding with the estimates $\rho_n^I$ for $n=1,2$ at all the three working temperatures. This is not very concerning, as the best estimate in this case is $\rho_3^I$ since it takes into account the highest order charge cumulant $\chi_8^I$. We observe that though the first zeros appear far away from $\rho_1^I$ and $\rho_2^I$, they are very near to coinciding with $\rho_3^I$ for all these three temperatures. 
Although the zeros start appearing beyond the estimate of $\rho_3^I$ at $157$ MeV, they are consistent and within errorbars of the individual estimates, are in good agreement with $\rho_3^I$ at $135$ and $176$ MeV.
Thus, the point of the first appearance of zeros of $|\LA e^{i\theta} \RA|$ procured using second and fourth order unbiased resummation approaches, provides commendable indications about the radius of convergence and the possible breakdown at least for eighth order Taylor series. This is somewhat promising, although the agreement with the same is not so well-defined at $157$ MeV. We may have to look out for some other indicators of radius of convergence at this temperature. It may also signify that maybe in order to capture the genuine behaviour of the Taylor series at $157$ MeV, it is important to go to even higher-than-eighth order calculations or it may also require the determination of unbiased phasefactor from a higher-than-sixth-order unbiased exponential resummation.

\begin{figure}[H]
    \centering
    \includegraphics[width=.325\textwidth]{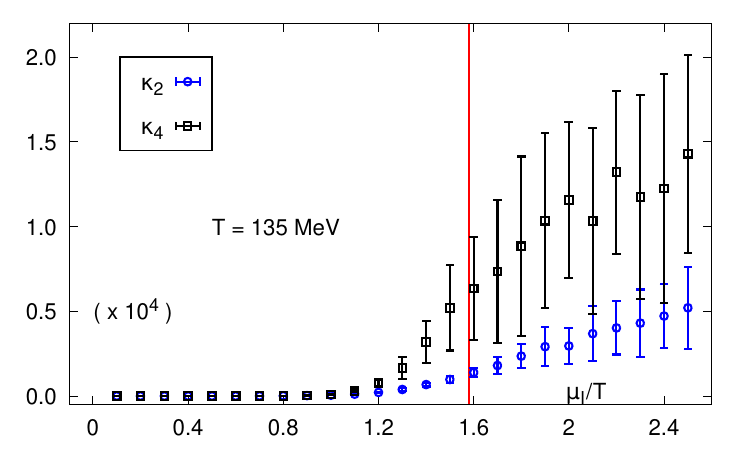}
    \includegraphics[width=.325\textwidth]{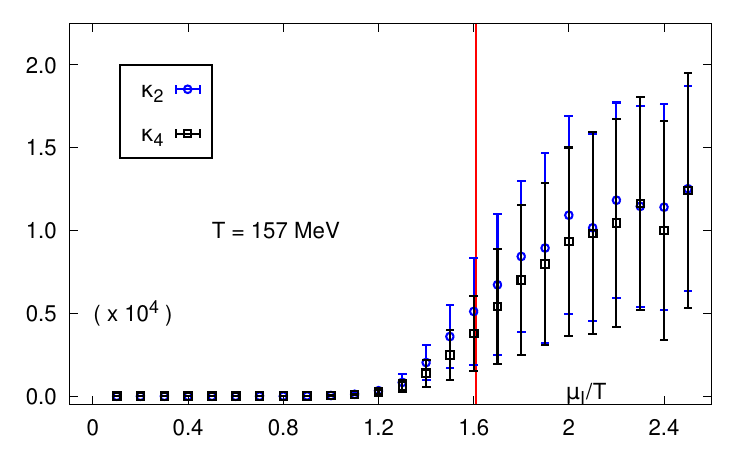}
    \includegraphics[width=.325\textwidth]{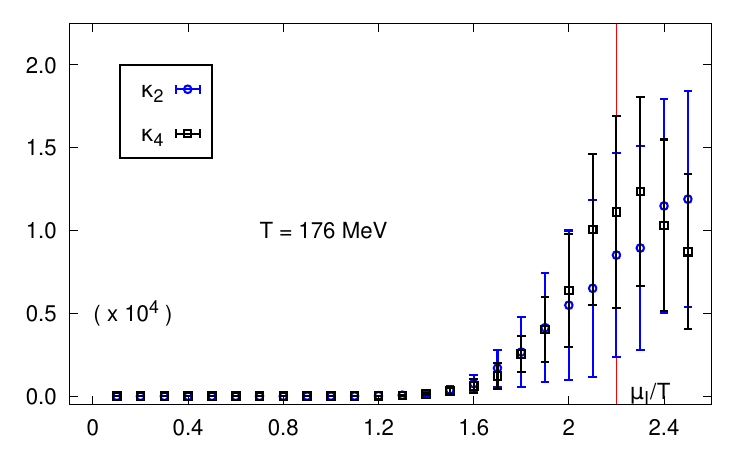} 
    
    \caption{Plots of second and fourth order kurtosis $\kappa_2$ and $\kappa_4$ as a function of $\muI/T$ for $135$ (left), $157$ (center) and $176$ MeV (right). The blue and black points illustrate the second and fourth order kurtosis and the red vertical line illustrates the point of first zero of $\LA e^{i\theta}\RA$ as shown above in Fig.\ref{fig:ROC 135,157,176_I all with phasefactor_2nd}.}
    \label{fig:135,157,176_I kurtosis}
\end{figure}

Lastly, we also present results manifesting the behaviour of kurtosis as a function of $\muI/T$. The kurtosis offers a quantitative measure of the overlap problem and its severity, which have been briefly discussed before. This overlap problem becomes visibly predominant for $\muI$ where there is no sign problem, and maybe the most possible reason for a breakdown. In Fig.\ref{fig:135,157,176_I kurtosis}, we observe that both the second and fourth order kurtosis exudes a monotonically increasing behaviour for higher values of $\muI$. This is expected since the extent of overlap between distributions generated at finite $\muI$ and zero $\muI$ reduces with increasing value of the finite $\muI$, and is evident from the higher values of the associated errorbars as seen in Fig.\ref{fig:135,157,176_I kurtosis}. The red vertical line depicts the appearance of first zero of $\LA e^{i\theta} \RA$, which coincides with the value of magenta line shown in Fig.\ref{fig:ROC 135,157,176_I all with phasefactor_2nd} as mentioned before. Fig.\ref{fig:135,157,176_I kurtosis} demonstrates that for all the three temperatures, the errorbars increases rapidly across this red line. This indicates that as the oscillations of the complex fermionic determinant for complex $\muI$ increases making $\LA e^{i\theta} \RA \approx 0$, the overlap problem rapidly increases. Apart from the increasing errorbars, this is also manifested by the non-monotonic behaviour of the second and fourth order kurtosis values. This is an encouraging sight as one can understand the severity of this overlap problem and associated breakdown by studying the behaviour of this newly proposed complex phasefactor and observing the value of $\muI$ at which its first zero manifests. 

% of this newly proposed complex phasefactor does offer for this overlap problem is consistent for all the working temperatures, to a good extent with the onset of the zero of the proposed complex phasefactor $\left|\LA e^{i\theta}\RA\right|$. Beyond the value of $\muI$ at which the first zero of $\left|\LA e^{i\theta}\RA\right|$ appears, the overlap problem becomes very severe and the errorbars also become extremely large along with a non-monotonic behaviour of the mean values of $\kappa_2$ and $\kappa_4$.

\section{Conclusions}
\label{sec:Conclusions}

In this paper, we have conducted a detailed comparative study between the estimates of higher order charge cumulants, namely $\chi_6$ and $\chi_8$ for both $\muB$ and $\muI$ which have been obtained using the approaches of Taylor expansion and the unbiased exponential resummation. The estimates of these cumulants are exact in Taylor expansion, in the sense that all the necessary and appropriate correlation functions have been included with proper estimation of their unbiased powers before averaging them over the working gauge ensemble. On the other hand, correlation functions only upto order $N$ are kept in $N^{th}$ order unbiased exponential resummation and it is only from these non-zero correlation functions in this approach that the respective charge cumulants of sixth and eighth orders are evaluated and estimated.

Although there still remains a lot to investigate and scope for further progress, it has been evident from the aforementioned results that for $\muB$, the resummed estimates of the sixth and eighth order charge cumulants seem to give excellent agreement with the corresponding Taylor estimates. More importantly, this commendable compatibility and agreement between the estimates of these two approaches is achieved for all the working temperatures from unbiased exponential resummation of fourth order onwards. However, this is not completely true for $\muI$ where we find that there are some discrepancies between the fourth order resummation estimate and the Taylor estimate for $\chi_8^I$, which are considerably reduced when one performs this resummation to sixth order in $\muI$. This is also the very reason why the isospin phasefactor has been calculated and procured from sixth order unbiased exponential resummation. A detailed study behind this discrepancy and other important features is certainly a work for future. We also have computed and compared the various estimates of radius of convergence between these two formalisms. WE have observed that except the second order resummation, the fourth and sixth order resummation estimates agree very well with the corresponding Taylor estimate for all the working temperatures for both $\muB$ and $\muI$. 

% Although a lot more needs to be done before confirming this statement, the current set of results presented in this paper seem to clearly imply that once the stochastic bias upto $\mathcal{O}(\mu^4)$ is eliminated by considering the necessary unbiased powers of $D_n$ upto $n=4$ using unbiased exponential resummation, the resummation results work very well and agree with the Taylor-ed estimates of charge cumulants of at least upto eighth orders. More importantly, this observation holds true more or less for all the three working temperatures in this paper, which are subtly selected so that they encompass a large part of the entire QCD phase diagram. This finding also remains true irrespective of whether or not, one calculates unbiased powers of the higher $D_n$ from $n=5$ onwards. So, this may possibly imply that owing to the ultraviolet divergences remaining upto $\mathcal{O}(\mu^4)$, it is only these first four correlation functions which impacts the resulting calculations maximally and therefore must be treated in an unbiased manner for procuring the behaviour of the true infinite Taylor series and for conducting a more transparent genuine probe into QCD at finite chemical potential.

We have also investigated the zeros of the gauge-ensemble average phasefactor and found that the zeros in case of $\muB$ provide promising indications of radius of convergence of Taylor series and subsequent breakdown. Apart from $\muB$, this also holds true to some extent for $\muI$ where we have proposed a new formula for a non-trivial phasefactor by considering complex $\muI$. Except $157$ MeV, we found that the first zeros of this newly formulated phasefactor align very well with the estimate of $\rho_3^I$ for $135$ and $176$ MeV and therefore can offer with some degree of reliability, promising indications about the onset of breakdown of calculations and also the estimate of radius of convergence of the associated Taylor series. At the end, we also demonstrated the nature of the overlap problem with $\muI$ and illustrated that the onset of the zeros of this newly proposed phasefactor is more or less consistent with this problem and this overlap problem becomes severely large, with excessively high errorbars and non-monotonic behaviour beyond the point of these first zeros, for all the three working temperatures.

\acknowledgments

 I sincerely acknowledge Prasad Hegde and Frithjof Karsch for highly useful discussions and stimulating constructive suggestions for this draft. I also sincerely thank all the other members of the HotQCD collaboration for their inputs and valuable insights, as well as for allowing me to use their data for the respective Taylor expansion calculations. The computations in this work have been performed on the GPU cluster at Bielefeld University, Germany. I also heartily thank the Bielefeld HPC.NRW team for their wholehearted support. 

%\vspace{.1cm}

 \bibliographystyle{JHEP}
 %\bibliography{main}
 
\providecommand{\href}[2]{#2}\begingroup\raggedright\endgroup

\vspace{1cm}

%% or
%% [B] Manual formatting (see below)
%% (i) We suggest to always provide author, title and journal data or doi:
%% in short all the informations that clearly identify a document.
%% (ii) please avoid comments such as "For a review'', "For some examples",
%% "and references therein" or move them in the text. In general, please leave only references in the bibliography and move all
%% accessory text in footnotes.
%% (iii) Also, please have only one work for each \bibitem.

% \begin{thebibliography}{99}

% \bibitem{a}
% Author,
% \emph{Title},
% \emph{J. Abbrev.} {\bf vol} (year) pg.

% \bibitem{b}
% Author,
% \emph{Title},
% arxiv:1234.5678.

% \bibitem{c}
% Author,
% \emph{Title},
% Publisher (year).

% \end{thebibliography}
\end{document}